# Zonal jets experiments in the gas giants' zonostrophic regime

D. Lemasquerier[a,*,1],  B. Favier[a] and  M. Le Bars[a]

[a]*Aix Marseille Univ, CNRS, Centrale Marseille, IRPHE, , Marseille, 13013, France*

ABSTRACT

Intense east-west winds called zonal jets are observed in the atmospheres of Jupiter and Saturn and extend in their deep interior. We present experimental results from a fully three-dimensional laboratory analog of deep gas giants zonal jets. We use a rapidly rotating deep cylindrical tank, filled with water, and forced by a small-scale hydraulic circulation at the bottom. A topographic $\beta$-effect is naturally present because of the curvature of the free surface. Instantaneous turbulent zonal jets spontaneously emerge from the small-scale forcing, equilibrate at large scale, and can contain up to 70% of the total kinetic energy of the flow once in a quasi-steady state. We show that the spectral properties of the experimental flows are consistent with the theoretical predictions in the zonostrophic turbulence regime, argued to be relevant to gas giants. This constitutes the first fully-experimental validation of the zonostrophic theory in a completely three-dimensional framework. Complementary, quasi-geostrophic (QG) simulations show that this result is not sensitive to the forcing scale. Next, we quantify the potential vorticity (PV) mixing. While PV staircasing should emerge in the asymptotic regime of the gas giants, only a moderate PV mixing occurs because of the strong forcing and dissipation, as confirmed by QG simulations at smaller Ekman number. We quantify the local PV mixing by measuring the equivalent of a Thorpe scale, and confirm that it can be used to estimate the upscale energy transfer rate of the flow, which otherwise needs to be estimated from a much more demanding spectral analysis.

## 1. Introduction

Zonal jets are east-west currents observed in a large variety of geophysical flows, from the atmospheres of gas giants (Vasavada and Showman, 2005) to terrestrial oceans (Maximenko, Bang and Sasaki, 2005) and atmospheres (Mitchell, Birner, Lapeyre, Nakamura, Read, Riviére, Sánchez-Lavega and Vallis, 2019). They are also expected to develop in buried fluid layers such as the outer liquid cores of telluric planets (Guervilly and Cardin, 2017) or the subsurface oceans of icy moons (Soderlund, Kalousová, Buffo, Glein, Goodman, Mitri, Patterson, Postberg, Rovira-Navarro, Rückriemen, Saur, Schmidt, Sotin, Spohn, Tobie, Van Hoolst, Vance and Vermeersen, 2020). The vast occurrence of zonal jets underlines their generic nature, which relies on three basic physical effects: the dominance of rotation on the flow, the presence of turbulent motions forced by various processes, and a $\beta$-effect which represents the variation of the Coriolis force with latitude. Apart from their clear geophysical relevance, zonal jets are outstanding fluid dynamical features where a turbulent flow self-organizes at large-scale. The properties of the associated statistically steady-state in terms of number, width, intensity and stability of the jets are described by several theories based on the phenomenology of turbulence (Vallis and Maltrud, 1993; Galperin, Sukoriansky, Dikovskaya, Read, Yamazaki and Wordsworth, 2006), potential vorticity mixing (Dritschel and McIntyre, 2008), wave-mean flow interaction (Quinn, Nazarenko, Connaughton, Gallagher and Hnat, 2019) or statistical physics (Bouchet and Venaille, 2012; Farrell and Ioannou, 2003), but a common and robust physical framework is still lacking.

In the present study, we specifically focus on the regime of turbulent zonal jets observed on the gas giants, Jupiter and Saturn, where the jets' dynamics is not complicated by topographic features or seasonal variations. Jupiter's zonal jets are famous because they are responsible for the banded appearance of the planet, through the spreading of clouds of ammonia and water ices. Jupiter's intense zonal winds have been measured by various spacecrafts using cloud tracking, from *Voyager* in 1979 to *Juno*, which is in orbit around Jupiter since 2016 (Bolton, Lunine, Stevenson, Connerney,

Levin, Owen, Bagenal, Gautier, Ingersoll, Orton, Guillot, Hubbard, Bloxham, Coradini, Stephens, Mokashi, Thorne and Thorpe, 2017). Direct measurements revealed the remarkable stability of these strong winds over decades (Tollefson, Wong, de Pater, Simon, Orton, Rogers, Atreya, Cosentino, Januszewski, Morales-Juberías et al., 2017), and allowed to demonstrate that they contain up to 90% of the kinetic energy of the turbulent flow, at least at the level of the cloud layer (Galperin, Young, Sukoriansky, Dikovskaya, Read, Lancaster and Armstrong, 2014b). Recently, the inversion of the gravity measurements of the Juno spacecraft showed that the Jovian jets extend down to about 3,000 kilometers beneath the clouds for Jupiter (Kaspi, Galanti, Hubbard, Stevenson, Bolton, Iess, Guillot, Bloxham, Connerney, Cao, Durante, Folkner, Helled, Ingersoll, Levin, Lunine, Miguel, Militzer, Parisi and Wahl, 2018; Guillot, Miguel, Militzer, Hubbard, Kaspi, Galanti, Cao, Helled, Wahl, Iess et al., 2018), and 9,000 kilometers for Saturn (Galanti, Kaspi, Miguel, Guillot, Durante, Racioppa and Iess, 2019) based on Cassini measurements. This result is consistent with independent constraints provided by the secular variation of Jupiter's magnetic field (Moore, Cao, Bloxham, Stevenson, Connerney and Bolton, 2019). Despite the wealth of new constraints provided by *Juno*, comprehensive idealized models still need to be developed to complement observations and better understand the jets origin, properties, and three-dimensional structure.

From the fluid dynamics point of view, gas giants exhibit very intense flows, characterized by a very large Reynolds number ($Re$), a vanishing Ekman number ($E$) and a small Rossby number ($Ro$). The Reynolds number quantifies the ratio of inertial to viscous forces, the Ekman number the ratio of viscous to Coriolis forces, and the Rossby number the ratio of inertial to Coriolis forces. Denoting $U$ a typical velocity scale, $H$ a typical flow length scale, $\nu$ the molecular viscosity of the fluid and $\Omega$ the planet's rotation rate, these non-dimensional parameters can be expressed as

$$Re \sim \frac{UH}{\nu}, \quad E \sim \frac{\nu}{\Omega H^2}, \quad Ro \sim \frac{U}{\Omega H}. \tag{1}$$

The aforementioned conditions hence mean that Jupiter's flows are both highly turbulent and rotationally constrained (see Table 1 and Fig.1). Because rotation is dominant, the turbulent motions are quasi two-dimensional (2D) and may bear an inverse cascade of kinetic energy (Young and Read, 2017; Siegelman, Klein, Ingersoll, Ewald, Young, Bracco, Mura, Adriani, Grassi, Plainaki and Sindoni, 2022) feeding large-scale features (vortices and jets). We denote $\epsilon$ the corresponding rate of energy transfer. The zonation of Jupiter's turbulence then arises because of the so-called $\beta$-effect, representing either the latitudinal variation of the projection of the planet's rotation vector on the local vertical (shallow, atmospheric $\beta$-effect), or the variation of the fluid height when measured parallel to the rotation axis (deep, topographic $\beta$-effect). Because of the $\beta$-effect and associated Rossby waves, the turbulent energy transfer to large scales becomes anisotropic and redirected towards zonal currents. Ultimately, the large scale drag halts the development of the inverse cascade (Sukoriansky, Galperin and Dikovskaya, 2002; Sukoriansky, Dikovskaya and Galperin, 2007). A fourth non-dimensional parameter, known as the zonostrophy index $R_\beta$ (Sukoriansky et al., 2007), is used as an indicator of the strength of the zonation of the turbulent flow. It is defined as the ratio between the Rhines scale (Rhines, 1975) $L_R$,

$$L_R \propto \left(\frac{2U}{\beta}\right)^{1/2}, \tag{2}$$

assumed to represent the scale of the jets, and the transitional scale (Pelinovsky, 1978; Vallis and Maltrud, 1993) $L_\beta$,

$$L_\beta \propto \left(\frac{\epsilon}{\beta^3}\right)^{1/5}, \tag{3}$$

which is the scale at which a turbulent eddy turnover time equates a Rossby wave period. The transitional scale is the smallest scale above which the turbulent flow becomes anisotropic due to the $\beta$-effect. The zonostrophy index is then

$$R_\beta \sim \frac{L_R}{L_\beta} \sim 2^{1/2} \left(\frac{U^5 \beta}{\epsilon^2}\right)^{1/10} \tag{4}$$

(Sukoriansky et al., 2007). The regime of strong and rectilinear jets – so-called *zonostrophic regime* – is obtained when the scale at which the eddies start being deformed by the $\beta$-effect is well separated from the scale of the real jets, i.e. for a large zonostrophy index. From 2D simulations on the sphere, Sukoriansky et al. (2007); Galperin, Sukoriansky and Dikovskaya (2010) show that the zonostrophic regime is observed when $R_\beta^* \gtrsim 2.5$ whereas the regime is friction-dominated when $R_\beta^* \lesssim 1.5$, as represented in Fig. 1(*b*). Here, $R_\beta^* = (C_Z/C_K)^{3/10} R_\beta \approx 0.5 R_\beta$, and $C_Z$ and $C_K$ are supposedly universal constants determined from spectral analysis (see section 4). Jovian mid-latitude jets have $R_\beta^* \approx 5$ and are expected to be in the regime of zonostrophic turbulence (Galperin et al., 2014b). On the contrary, Earth's oceans are under the threshold with $R_\beta^* \approx 1.5$ (Galperin, Sukoriansky, Young, Chemke, Kaspi, Read and Dikovskaya, 2019). Consistently, Jovian jets are strong, instantaneous and contain most of the kinetic energy of the flow (Galperin et al., 2014b) whereas oceanic jets are weak and meandering, hence a careful time-averaging is required to reveal them (Maximenko et al., 2005).

**Table 1**
Non-dimensional parameters and typical values. The Ekman number, $E$, Reynolds number $Re$ and zonostrophy index $R_\beta$ are independent, the Rossby number $Ro$ is indicated for completeness but can be expressed as $Ro = Re \times E$. For the experiments and QG simulations, assuming dissipation in the Ekman boundary layers (equation (7)), these parameters can be evaluated either using the fluid height $h_0$ (first column) or the upscale energy transfer rate $\epsilon$ (second column). $u_{rms}$ is the total and time-averaged rms velocity measured once in statistically steady-state. This is relevant only for the experiments and QG simulations, not the planetary flows. The planetary parameters used to compute the non-dimensional parameters for the gas giants and Earth's atmosphere and oceans are provided in Appendix B.

| | $(u_{rms}, h_0, \Omega, \nu, \beta)$ | $(u_{rms}, \epsilon, \Omega, \nu, \beta)$ | Jupiter | Saturn | Earth's atmosphere | Earth's oceans | Experiment |
|---|---|---|---|---|---|---|---|
| $E$ | $\dfrac{\nu}{\Omega h_0^2}$ | $\dfrac{4\epsilon^2}{\Omega^2 u_{rms}^4}$ | $4 \times 10^{-16}$ | $4 \times 10^{-17}$ | $2 \times 10^{-9}$ | $1 \times 10^{-8}$ | $3 \times 10^{-7}$ |
| $Re$ | $\dfrac{u_{rms} h_0}{\nu}$ | $\dfrac{u_{rms}^3 \Omega^{1/2}}{2\epsilon \nu^{1/2}}$ | $3 \times 10^{14}$ | $9 \times 10^{14}$ | $3 \times 10^9$ | $1 \times 10^8$ | $1 \times 10^4$ |
| $Ro$ | $\dfrac{u_{rms}}{\Omega h_0}$ | $\dfrac{2\epsilon}{\Omega^{3/2}\nu^{1/2}u_{rms}}$ | $1 \times 10^{-1}$ | $4 \times 10^{-2}$ | 6 | 1 | $3 \times 10^{-3}$ |
| $R_\beta^*$ | $\left(\dfrac{C_Z}{C_K}\right)^{3/10}\left(\dfrac{2^7 \beta u_{rms} h_0^2}{\nu \Omega}\right)^{1/10}$ | $\left(\dfrac{C_Z}{C_K}\right)^{3/10}\left(\dfrac{2^5 u_{rms}^5 \beta}{\epsilon^2}\right)^{1/10}$ | $5^a$ | $5.3^a$ | $1.6^a$ | $1.4^a$ | $\gtrsim 2.5$ |

$^a$see Table 13.1 in Galperin and Read (2019)

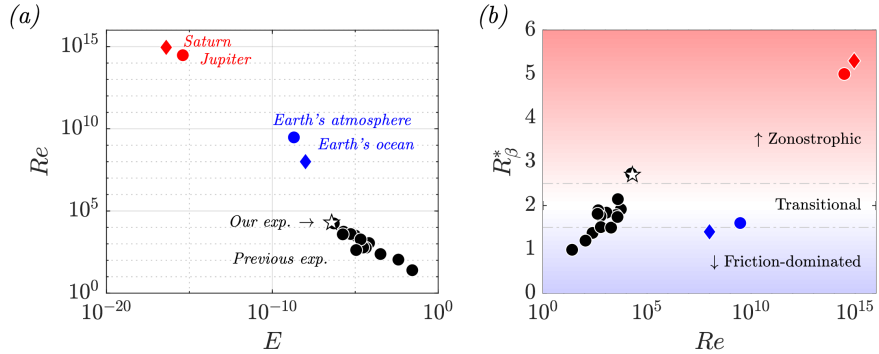

**Figure 1:** Location of previous zonal jets laboratory experiments in the non-dimensional parameter space $(Re, E, R_\beta^*)$ (Reynolds number, Ekman number, and zonostrophy index). For the sake of comparison, we used similar definitions for all the experimental studies (first column of Table 1). The data used to plot this figure are provided in Appendix B. We recall that $R_\beta^* \approx 0.5R_\beta$ and takes into account the spectral prefactor $(C_Z/C_K)^{3/10}$ (see section 4). The horizontal dashed lines represent the threshold in $R_\beta^*$ between the different regimes determined by Sukoriansky et al. (2007); Galperin et al. (2010) .

Modeling turbulent zonal jets in regimes relevant to gas giants requires to fulfill these four dynamical constraints simultaneously ($Re \gg 1, E \ll 1, Ro \lesssim 1, R_\beta^* \gtrsim 2.5$). This is particularly challenging, both from the numerical and experimental point of view because of computational and technical constraints. This regime can only be approached by simplified models relying on physical approximations whose relevance needs to be systematically addressed. For instance, current 3D numerical models cannot reach the asymptotic Reynolds and Ekman numbers of gas giants flows, and when trying to do so, it becomes extremely costly to simulate the dynamics over the very long radiative and frictional timescales, necessary for the system to equilibrate (see e.g. Heimpel, Gastine and Wicht, 2016; Gastine and Wicht, 2021). On the contrary, experiments usually allow to reach more extreme regimes than numerical models, and study the dynamics of real, fully nonlinear and developed flows in a statistically steady-state, even if they cannot reproduce the very small viscous effects of Jovian flows. Of course, experiments come with their own limitations, such as necessary physical boundaries, partial measurements, and the difficulty in incorporating magneto-hydrodynamical or compressibility effects.

In a previous paper (Lemasquerier, Favier and Le Bars, 2021), we described a new experimental setup following on from Cabanes, Aurnou, Favier and Le Bars (2017) where zonal jets spontaneously emerge from a mechanically-forced turbulent flow in a rapidly rotating water tank. This setup is a one layer system, with a deep water height, and hence the zonal jets are deep and barotropic. Here, we would like to stress out that a barotropic (one layer) experimental setup is still relevant to model *a priori* strongly baroclinic systems, such as oceans and atmospheres. Indeed, planetary, quasi-geostrophic flows can undergo a phenomenon of *barotropization*, and the barotropic modes can end up containing a significant fraction of the energy, even if the energy feeding these modes initially comes from baroclinic modes and instabilities (Charney, 1971; Rhines, 1977; Salmon, 1978) (see section 6.3 for further discussion). In the experiment of Lemasquerier et al. (2021), the zonal jets emerge due to the radiation of Rossby waves and the associated transport and deposition of momentum. In addition, we observed a transition between two regimes of multiple zonal jets. We showed that the transition arises from a Rossby waves resonance due to their advection by the background zonal flow. We showed that the first regime (hereafter Regime I), where the jets are individual and locally forced, may be relevant for jets in the terrestrial oceans. The second regime (hereafter Regime II) is accompanied by a coarsening and an intensification of the zonal jets, and is relevant for the gas giants. In the present study, we focus on the saturated turbulent and statistically steady state obtained in Regime II, far from the transition, and we compare our experimental results with two theories which aim to explain zonal jets properties at a global scale.

The first theory on which we focus is the so-called theory of zonostrophic turbulence (see Galperin et al., 2019). In this framework, zonal jets are described as emerging from the anisotropisation of two-dimensional turbulence in the presence of a $\beta$-effect. However, this description may only be relevant when the four aforementioned dynamical constraints are fulfilled, and when sufficient scale separations are achieved such that turbulent inertial ranges can exist. Following Cabanes et al. (2017), our experimental setup is specifically designed to favor a large $R_\beta$, fast flows ($Re \gg 1$), but still dominated by rotation ($Ro \ll 1$), and small viscous dissipation ($E \ll 1$), thus getting closer to the regime observed on gas giants. As shown in Fig.1(*b*), previous experimental studies lied in the range $R_\beta^* \in [1, 2]$ and the observed flows were not in the zonostrophic regime. Let us stress out that, because of the definition of the zonostrophy index and the power $1/10$ (see Table 1), if all the other parameters are unchanged, a rms velocity ten times larger (and hence a Reynolds ten times larger) only increases the zonostrophy index by a factor $10^{1/10} \sim 1.26$! Here, we explore more extreme regimes, including $R_\beta^* \gtrsim 2.5$. The experimental measurements can hence be compared with predictions in the so-called zonostrophic regime relevant to gas giants. The spectral analysis performed in (Cabanes et al., 2017) on a previous version of the experiment supports the idea that predictions from zonostrophic turbulence are retrieved experimentally. Kinetic energy spectra were however computed for the zonal flow only and not the fluctuations, due to the use of particle tracking velocimetry. With the new setup (Lemasquerier et al., 2021), the temporal and spatial resolution of our PIV measurements allows us to quantify turbulence statistics for both the zonal and fluctuating components of the flow, and to complete the analysis in the framework of zonostrophic turbulence.

The second theoretical framework to which we compare our experimental results is potential vorticity (PV) mixing (Dunkerton and Scott, 2008; McIntyre, 2008; Dritschel and McIntyre, 2008; Scott and Tissier, 2012). In our case of a barotropic fluid with varying height, the PV is $q = (\zeta + f)/h$, where $\zeta$ is the relative vorticity, $f = 2\Omega$ is the "planetary" vorticity ($\Omega$ is the rotation rate), and $h$ is the fluid height. In the framework of PV mixing, zonal jets are described as emerging from the mixing of the scalar field $q$, which is materially conserved in the absence of dissipation. Retrograde jets correspond to well-mixed regions of PV, whereas prograde jets correspond to steep gradients of PV, thus defining a staircase-like profile in the radial direction if multiple jets are present. The mechanism of staircase formation was described in Dritschel and McIntyre (2008) by analogy with the mixing of a stratified fluid, the so-called "Phillips-effect" (Phillips, 1972). The idea is that of a feedback mechanism: stratification is reduced in the mixed regions, the restoring force at the origin of gravity waves (buoyancy) is thus weakened, which allows for further mixing. On the contrary, where interfaces are forming, steeper stratification locally develop, and gravity waves are enhanced, thus inhibiting mixing across the interface. For a PV gradient instead of a density gradient, the mechanism is analogous but invokes Rossby waves whose propagation depends on how steep the local PV gradient is. That being said, subtleties arise from the fact that PV is different from density since there is a direct relationship between the flow dynamics and PV through the vorticity.

Galperin, Hoemann, Espa and Nitto (2014a) proposed to use the analogy between PV and density staircases to quantify PV mixing, through the method of PV monotonizing. The idea is to use local and instantaneous PV (or density) profiles to quantify the turbulent overturn by measuring the equivalent of a Thorpe scale (Thorpe, 2005; Gargett and Garner, 2008). The initial PV (density) profile, due to background rotation and $\beta$-effect, is monotonous and decreasing with radius. Locally, turbulent eddies can cause overturns and bring fluid parcels with higher PV (higher density) outward (above) compared to a fluid parcel with smaller PV (smaller density); an unstable configuration. By sorting the non-monotonous PV profile into a monotonous one, one can define a Thorpe scale $L_T$, which is the root-mean-squared displacement of the fluid parcels needed to bring them back to a stable position. Galperin et al. (2014a) show experimentally that similarly to stratified flows where $L_T$ is close to the Ozmidov scale (Ozmidov, 1965; Thorpe and Deacon, 1977), which is the scale at which turbulent eddies feel the stratification, $L_T$ is commensurate with the transitional scale, $L_\beta$, at which turbulent eddies feel the $\beta$-effect. If a robust relationship exists between the Thorpe and transitional scales, then sorting PV would constitute a powerful tool to estimate $L_\beta$ and indirectly, the turbulence intensity. The only requirement would then be to measure instantaneous profiles of potential vorticity of the flow (see for instance Cabanes, Espa, Galperin, Young and Read, 2020, for an application to Jupiter and Saturn). However, the three experiments used in (Galperin et al., 2014a) are far from the zonostrophic regime: the forcing, which is performed using magnets and a saline solution, accelerates a westward zonal flow locally but also directly (all the magnets are aligned and have the same polarity). To what extent the scaling between the Thorpe and the transitional scale holds in our experiments where the jets are strongly turbulent and self-developed from a fluctuating forcing with no azimuthal mean, thus remains to be addressed.

The present paper is organized as follows. In §2, we briefly recall the main characteristics of the experimental setup, and we present the numerical quasi-geostrophic model which is used to complement the experimental results. In §3, we describe the fully developed flows obtained in both the experiments and simulations. In §4, we compute kinetic energy spectra to compare with zonostrophic turbulence predictions and we measure a turbulent energy dissipation rate from the spectra. QG simulations are used to extend the experimental results to smaller scales and hence better scale separation. In §5, we address the question of potential vorticity mixing. At a global scale, we quantify PV staircaising and extend our experimental results with simulations performed at a smaller Ekman number. At a local scale, we compute the Thorpe scale associated with PV mixing, which provides a second independent way of estimating the upscale energy transfer rate. We conclude and discuss our results in §6.

## 2. Experimental and numerical methods

### 2.1. Experimental set-up

#### 2.1.1. Description of the setup

The experimental set-up is sketched in Fig.2(a) and is extensively described in Lemasquerier et al. (2021) and Lemasquerier (2021). For consistency, we summarize here its principal features. It consists in a rotating cylindrical water tank, of radius $R = 0.49$ m, filled with about 600L of water. The fast rotation of the tank induces a centrifugal force leading to a strongly deformed, paraboloidal free-surface. Since the fluid height varies with radius, the relative vorticity of a fluid column moving radially outward (stretching) or inward (squeezing) is respectively increased or decreased. This effect is the so-called topographic $\beta$-effect and is responsible for introducing anisotropy in the system in the form of zonation along the azimuthal direction. Here, the topographic $\beta$ parameter can be written as

$$\beta = -\frac{f}{h}\frac{\mathrm{d}h}{\mathrm{d}\rho}, \tag{5}$$

where $\rho$ is the cylindrical radius, $h(\rho)$ is the total fluid height and $f = 2\Omega$ is the Coriolis parameter with $\Omega$ the rotation rate. The vast majority of theoretical developments are performed in the $\beta$-plane approximation, where $\beta$ is spatially homogeneous – for planetary applications, this amounts to a local approach where the global variations of $\beta$ are neglected. Equation (5) shows that for the topographic $\beta$-effect to be uniform over the domain, the fluid height should vary exponentially with radius. To achieve this, we compensate the unalterable paraboloidal shape of the free surface using a curved bottom plate placed inside of the tank (Fig.2(c)). The total fluid height $h$ above the bottom plate is the difference between the free-surface altitude $h_p$ and the bottom topography altitude $h_b$. The bottom topography has been chosen such that, in solid body rotation at a fixed rate $\Omega$, the total water height as a function of the cylindrical radius $\rho$ is

$$h(\rho) = \underbrace{h_0 + \frac{\Omega^2}{2g}\left(\rho^2 - \frac{R^2}{2}\right)}_{h_p(\rho)} - h_b(\rho) = h_{\min}\exp\left(-\frac{\beta}{2\Omega}\rho\right), \tag{6}$$

where $g$ is the gravitational acceleration, $R$ is the tank radius, $h_0$ the fluid height at rest, $h_{\min}$ the minimum fluid height once in rotation and $\beta < 0$ the spatially uniform $\beta$-parameter (eq.(5)). We worked with $\Omega = 75$ RPM (rotation frequency of 1.25 Hz), $h_0 = 0.58$ m, $h_{\min} = 0.20$ m and $\beta = -50.1$ m$^{-1}$s$^{-1}$. Note that our experimental set-up was designed to work at a single rotation rate of 75 RPM at which the $\beta$-effect is then uniform across the tank. However, we performed few experiments at rotation rates of 60 and 80 RPM to investigate the consequences of a modified $\beta$-effect. At these rotation rates the $\beta$-effect is no more uniform but nevertheless varies in a limited range ($|\beta_{80}| \in [57, 73]$ m$^{-1}$ s$^{-1}$ with a mean at 65.5 m$^{-1}$ s$^{-1}$ and $|\beta_{60}| \in [20, 30]$ m$^{-1}$ s$^{-1}$ with a mean at 22.8 m$^{-1}$ s$^{-1}$).

Once solid-body rotation is reached, the flow is forced by circulating the tank's water through 128 holes drilled into the curved bottom plate (see Fig.2(a) and (c)). The holes are distributed on a polar lattice composed of six rings each connected to a separate pump. Half of the holes are inlets, and the other half are outlets, with as many inlets as outlets along each ring. Because of the dominant rotation, this forcing generates an array of cyclonic and anticyclonic vortices, which subsequently break into turbulence. The zonal average of the forcing is zero by construction, both along each ring and in total, meaning that we do no directly accelerate the zonal flow. Additional details on the experiment can be found in Lemasquerier et al. (2021) and D. Lemasquerier's thesis manuscript (Lemasquerier, 2021). Two different sets of pumps were used to reach different forcing amplitudes. They are described in Appendix B of Lemasquerier (2021).

To measure velocity fields, time-resolving particle image velocimetry (PIV) measurements are performed on a horizontal plane. A green horizontal laser sheet (532 nm) crosses the water layer 11 cm above the bottom plate (9 cm below the center of the paraboloid). The water is seeded with fluorescent red polyethylene particles and their motion is tracked using a top-view camera placed above the tank and embarked in solid-body rotation with the tank and the laser (Fig.2(a)). The particles emit an orange light (607 nm) so that using a high-pass filter on the lens allows to filter out the green laser reflections on the free-surface and tank sides, leading to a better image quality and hence better PIV measurements. The images are acquired using DANTEC's software DynamicStudio. Optical distortion induced by the paraboloidal free-surface is corrected on DynamicStudio using a preliminary calibration performed by imaging a home-made calibration target with a precise dot pattern (Lemasquerier, 2021). A movie of the particles motion during a typical experiment is available as supplementary movie 1 in Lemasquerier et al. (2021). The velocity fields are deduced from these images using the MATLAB program DPIVSoft developed by Meunier and Leweke (2003). We consider 32×32 pixels boxes on 1900×1900 pixels images and obtain 100×100 velocity vector fields (40% overlap between the boxes). Note that due to the refraction of the laser plane by the tank sides, there are two shadow zones where measurements are not possible (see the grey areas in Fig.5(a)).

Note that the present setup was built from scratch but inspired from the setup used in Cabanes et al. (2017). In addition to the time-resolved PIV measurements mentioned above, the main improvements of our setup compared to the one of Cabanes et al. (2017) include:

– A spatially-uniform $\beta$-effect to be able to compare with theories derived in the $\beta$-plane approximation.

– A control of the spatial distribution of the forcing. In Cabanes et al. (2017), all the injection and suction points were linked to the same pump. Here, the forcing pattern is polar such that we can control independently the six forcing rings using six different pumps and modulate the intensity of the forcing with radius.

– An increase in the number of forcing points from 64 to 128 in order to decrease the forcing scale and have a better scale separation with the transitional scale $L_\beta$.

### 2.1.2. Non-dimensional parameters

The experimental flows we consider are defined by five dimensional parameters, the rotation rate, $\Omega$, the mean fluid height, $h_0$, the molecular viscosity, $\nu$, the initial gradient of potential vorticity $\beta$ and the root-mean-squared (rms) velocity, $u_{\rm rms}$. Here, $u_{\rm rms}$ is an output parameter measured as the total rms velocity once in steady state. It is linearly related to the forcing amplitude, which is an input parameter set by choosing the power of the pumps (see Lemasquerier et al., 2021). With two dimensions (length and time), our experiments can be characterized by three independent non-dimensional parameters, which we chose to be the Ekman number, $E$, the Reynolds number, $Re$ and the zonostrophy index $R_\beta$, defined in the introduction, where we chose $u_{\text{rms}}$ and $h_0$ as the typical fluid velocity and length scales $U$ and $H$, respectively.

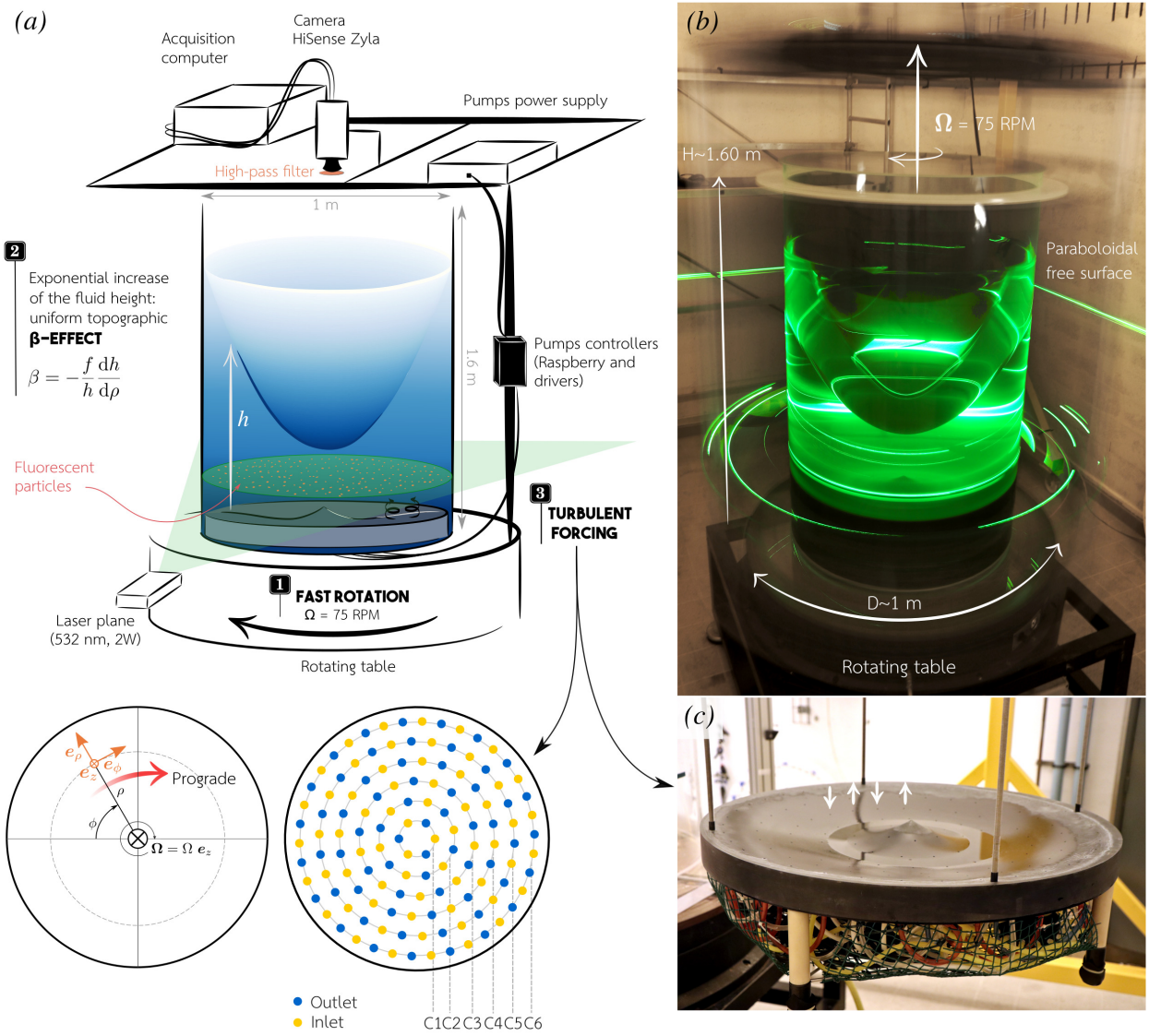

**Figure 2:** Experimental set-up. (a) Schematic of the experimental setup. The forcing pattern is sketched at the bottom: each ring C1–C6 is controlled by an independent pump. (b) Picture of the set-up once in solid body rotation. (c) Picture of the bottom forcing plate designed to achieve a uniform topographic $\beta$-effect.

In our laboratory experiments (but not in planetary flows), a consequence of dominant rotation is that one can assume that dissipation mainly occurs in the Ekman boundary layer forming along the frictional bottom plate. In such case, the upscale energy transfer $\epsilon$ can be expressed as a function of the Ekman spin-down timescale $\tau_E = \Omega^{-1} E^{-1/2}$:

$$\epsilon \sim \frac{u_{\text{rms}}^2}{2\tau_E} = \frac{u_{\text{rms}}^2 (\nu\Omega)^{1/2}}{2h_0}. \tag{7}$$

We will see later that $\epsilon$ can actually be measured, and agree relatively well with this estimate (section 4). In the remaining of the paper, and as indicated in Table 2, we make the difference between the *estimated* energy transfer rate, $\epsilon^E$, the transfer rate *measured* from the spectra, $\epsilon^S$, and the transfer rate measured from local potential vorticity mixing, $\epsilon^T$. Let us define a frictional lengthscale $L_E$ as $L_E \sim 2u_{\mathrm{rms}}\tau_E = u_{\mathrm{rms}}^3 \epsilon^{-1}$. The zonostrophy index can then be alternatively expressed as:

$$R_\beta = \frac{L_R}{L_\beta} = 2^{1/2}\left(\frac{L_E}{L_R}\right)^{1/5} = 2^{1/2}\left(\frac{L_E}{L_\beta}\right)^{1/6}. \qquad (8)$$

These expressions show that $R_\beta$ simultaneously compares the transitional scale, the Rhines scale, and the friction scale:

- if $R_\beta < 1$ ($L_E < L_R < L_\beta$), then the $\beta$-effect is weak and the scale of the flow is defined by the frictional scale $L_E$ which is then smaller than $L_R$ and $L_\beta$. In other words, since the frictional scale is smaller than the transitional scale, the energy is dissipated before the flow is fully affected by the $\beta$-effect. In that case, one would expect the flow to remain nearly isotropic and show features of the classical Kolmogorov-Batchelor-Kraichnan (KBK) turbulence (see Boffetta and Ecke, 2012, and references therein).

- if $R_\beta > 1$ ($L_\beta < L_R < L_E$), then the $\beta$-effect is strong, and the turbulence is anisotropic since its equilibrium scale is larger than the transitional scale $L_\beta$. The large scales are then limited at a value less than $L_E$.

Table 1 provides the expression of the non-dimensional parameters of the problem ($Re, E, R_\beta$) as a function of ($u_{\mathrm{rms}}, h_0, \Omega, \nu, \beta$) or alternatively ($u_{\mathrm{rms}}, \epsilon, \Omega, \nu, \beta$), where $\epsilon$ is approximated by equation (7). As previously mentioned, to model gas giants flows, both $Ro$ should be small and $Re$ should be large, which can only be achieved with an asymptotically small $E$, hence the need for a large tank, rotating fast. In our experiment, these constraints are fulfilled, as indicated by the values reported in Table 1. Note that our Ekman number is much higher than that of Jupiter, and our Reynolds is much smaller, but these discrepancies compensate in the Rossby number which has the good order of magnitude. Physically, this means that we have the good ratio of inertial to Coriolis forces, but our experimental flow is overly damped by bulk viscous effects which are vanishingly small on Jupiter. We are nevertheless in the relevant regime, turbulent while still dominated by rotation ($Re \gg 1$, $Ro \ll 1$).

**Table 2**
Symbols and definitions used in the paper

| Symbol | Expression | Definition |
|---|---|---|
| $k_R$ | $(\beta/2u_{\mathrm{rms}})^{1/2}$ | Rhines wavenumber |
| $\epsilon^E$ | $(u_{\mathrm{rms}}^2/2)\Omega E^{1/2}$ | Inverse energy transfer rate estimated a priori assuming Ekman friction only |
| $\epsilon^S$ | fit on residual spectra | Inverse energy transfer rate measured on the spectra using Eq.(20) |
| $\epsilon^T$ | $(k_\beta^T)^5\beta^3$ | Inverse energy transfer rate deduced from the Thorpe scale |
| $k_\beta^E$ | $(\beta^3/\epsilon^E)^{1/5}$ | Transitional wavenumber estimated using $\epsilon^E$ |
| $k_\beta^\epsilon$ | $(\beta^3/\epsilon^S)^{1/5}$ | Transitional wavenumber estimated using $\epsilon^S$ |
| $k_\beta^S$ | $(C_Z/C_K)^{3/10}(\beta^3/\epsilon^S)^{1/5}$ | Transitional wavenumber estimated on the spectral slopes intersection |
| $R_\beta^E$ | $k_\beta^E/k_R$ | Zonostrophy index estimated using $k_\beta^E$ |
| $R_\beta^\epsilon$ | $k_\beta^\epsilon/k_R$ | Zonostrophy index based on $k_\beta^\epsilon$ |
| $R_\beta^S$ | $k_\beta^S/k_R$ | Zonostrophy index based on $k_\beta^S$ |

### 2.1.3. List of experiments

In Table 3, we report the physical and non-dimensional parameters of the experiments that are used in the following of the paper. Because of experimental constraints, it is not possible to vary independently the three non-dimensional parameters $E$, $Re$ and $R_\beta$. First, the essential control parameter that we have is the forcing intensity (i.e. indirectly $u_{\mathrm{rms}}$), which allows us to explore significantly different Reynolds numbers. However, increasing $u_{\mathrm{rms}}$ also increases $R_\beta$ by decreasing the relative importance of large-scale drag. Second, we cannot explore a large range of Ekman numbers because the rotation rate of the experiment is fixed, as mentioned above. Third, for the three experiments performed at a different rotation rate, both $E$ and $R_\beta$ are modified since changing the rotation rate also modifies the $\beta$-effect by changing the fluid height.

**Table 3**
Parameters of the experiments and QG simulations used in the present paper. $u_{rms}$ is averaged over the last 500 rotation times of each experiment (typically between $t = 2500\ t_R$ and $t = 3000\ t_R$). $\epsilon^E$ is an estimate of the upscale energy transfer rate based on dissipation in the Ekman boundary layers (equation (7)). The Ekman number, $E$, the Reynolds number $Re$ and the zonostrophy index $R_\beta$ are defined in Table 1. The superscript $\cdot^E$ indicates that $R_\beta^E$ is estimated *a priori* using $\epsilon^E$ (see Table 2).

| Label | $\beta$ ($m^{-1}\,s^{-1}$) | $u_{rms}$ ($cm\,s^{-1}$) | $\epsilon^E$ ($m^2\,s^{-3}$) | $E$ ($\times 10^{-7}$) | $Re$ ($\times 10^3$) | $Ro$ ($\times 10^{-3}$) | $R_\beta^E$ |
|---|---|---|---|---|---|---|---|
| A | 50.1 | 3.44 | $28.7 \times 10^{-7}$ | 3.78 | 20.0 | 7.57 | 4.98 |
| B | 50.1 | 2.73 | $18.1 \times 10^{-7}$ | 3.78 | 15.9 | 6.00 | 4.86 |
| C | 65.5 | 3.23 | $26.1 \times 10^{-7}$ | 3.55 | 18.8 | 6.55 | 5.04 |
| D | 22.8 | 3.60 | $28.0 \times 10^{-7}$ | 4.73 | 20.9 | 9.87 | 4.71 |
| E | 50.1 | 3.01 | $21.8 \times 10^{-7}$ | 3.78 | 17.4 | 6.60 | 4.91 |
| F | 50.1 | 3.36 | $27.2 \times 10^{-7}$ | 3.78 | 19.5 | 7.37 | 4.96 |
| G | 50.1 | 3.32 | $26.6 \times 10^{-7}$ | 3.78 | 19.2 | 7.28 | 4.96 |
| H | 50.1 | 1.64 | $6.46 \times 10^{-7}$ | 3.78 | 9.48 | 3.59 | 4.62 |
| I | 50.1 | 1.85 | $8.25 \times 10^{-7}$ | 3.78 | 10.7 | 4.05 | 4.68 |
| J | 50.1 | 1.88 | $8.51 \times 10^{-7}$ | 3.78 | 10.9 | 4.12 | 4.69 |
| K | 22.8 | 1.64 | $5.82 \times 10^{-7}$ | 4.73 | 9.52 | 4.50 | 4.36 |
| L | 50.1 | 1.19 | $3.40 \times 10^{-7}$ | 3.78 | 6.88 | 2.60 | 4.48 |
| M | 50.1 | 1.26 | $3.84 \times 10^{-7}$ | 3.78 | 7.31 | 2.77 | 4.50 |
| N | 50.1 | 0.92 | $2.07 \times 10^{-7}$ | 3.78 | 5.36 | 2.03 | 4.37 |
| QG1a | 50.1 | 2.21 | $6.90 \times 10^{-7}$ | 1.27 | 22.2 | 2.82 | 5.32 |
| QG1b | 50.1 | 3.06 | $9.00 \times 10^{-7}$ | 0.60 | 44.5 | 2.67 | 4.93 |
| QG1c | 50.1 | 2.96 | $4.88 \times 10^{-7}$ | 0.20 | 74.8 | 1.50 | 6.59 |
| QG2 | 50.1 | 2.68 | $10.1 \times 10^{-7}$ | 1.27 | 26.8 | 3.42 | 5.42 |
| QG3 | 50.1 | 2.40 | $5.67 \times 10^{-7}$ | 0.63 | 34.3 | 2.14 | 5.76 |
| QG4 | 50.1 | 2.45 | $4.92 \times 10^{-7}$ | 0.42 | 43.3 | 1.81 | 6.01 |

## 2.2. Quasi-geostrophic numerical model

To complement and strengthen the experimental results, we perform numerical simulations. The goal is in particular to explore the sensitivity of the results to the forcing scale and pattern, which is difficult to do experimentally. Indeed, it would require to design and build a new bottom plate for each forcing pattern, and the numbers of inlets and outlets (128) would be difficult to increase significantly. To numerically model the experiment, 3D direct numerical simulations (DNS) are not adapted neither as they are very computationally demanding, in particular because the simulation has to resolve a turbulent flow with both very large scale structures (the jets) and thin Ekman boundary layers that are essential for the long term dynamics (see Cabanes et al., 2017). Instead, we model the experiment using a quasi-geostrophic (QG), two-dimensional numerical model. QG models take advantage of the fast background rotation, or equivalently the small Rossby number of the system: since the geostrophic balance dominates the experimental flow, the flow is quasi two-dimensional. The curvature of the free-surface as well as the friction over the bottom (Ekman pumping) induce three-dimensional effects, but their weakness allows their incorporation into quasi-two-dimensional physical models. With a QG model, we can exactly match the experimental conditions in terms of Reynolds, Rossby and Ekman numbers at a moderate computational cost, such that we can run several simulations over time scales comparable with that of the experiments (thousands of rotation times, or equivalently tens of frictional time scales). This would have been inconceivable using 3D direct numerical simulations.

In Appendix C, we derive the QG equations used in our numerical code. The difference with the conventional QG model is that, following Sansón and Van Heijst (2000, 2002) we retain higher order, non-linear terms for the Ekman pumping and $\beta$-effect. We consider a polar domain delimited by the outer and inner radii $r_o = R$ and $r_i = 0.05R$, where $R$ is the total radius. The inner boundary, absent in the experiment, is introduced to avoid the coordinate singularity at the center of the domain. We use the polar coordinates $(\rho, \phi)$ and denote $(\mathbf{e}_\rho, \mathbf{e}_\phi)$ the associated unit vectors (Fig.2). We solve separately the axisymmetric (zonal flow) and non-axisymmetric motions (see e.g. Aubert, Gillet and Cardin, 2003). To this end, we perform a Reynolds decomposition of the velocity field into an azimuthal average plus some fluctuations, denoted with a prime:

$$\langle X \rangle_\phi = \frac{1}{2\pi} \int_0^{2\pi} X \, d\phi, \tag{9}$$

$$u_\phi = \langle u_\phi \rangle_\phi + u'_\phi = U_\phi(\rho, t) + u'_\phi(\rho, \phi, t) \tag{10}$$

$$u_\rho = \langle u_\rho \rangle_\phi + u'_\rho = U_\rho(\rho, t) + u'_\rho(\rho, \phi, t), \tag{11}$$

$$\zeta = \langle \zeta \rangle_\phi + \zeta' = \partial_\rho(\rho U_\phi)/\rho + (\partial_\rho(\rho u'_\phi) - \partial_\phi u'_\rho)/\rho \tag{12}$$

In the following, all the variables denoted with a tilde are non-dimensional using $1/f$ as the time-scale, and the radius of the domain $R$ as the length-scale. For the fluctuations (non-axisymmetric modes), we solve the 2D barotropic vorticity equation and associated Poisson equation for the streamfunction:

$$\frac{\partial \tilde{\zeta}}{\partial \tilde{t}} + \mathcal{J}(\tilde{q}, \tilde{\psi}) - \frac{E_R^{1/2}}{2\tilde{h}} \tilde{\nabla} \tilde{\psi} \cdot \tilde{\nabla} \tilde{q} = \frac{E_R}{2} \tilde{\nabla}^2 \tilde{\zeta} - \frac{E_R^{1/2}}{2\tilde{h}} \tilde{\zeta}(\tilde{\zeta} + 1) + \tilde{F}, \tag{13}$$

$$\tilde{\zeta} = -\frac{1}{\tilde{h}} \tilde{\nabla}^2 \tilde{\psi} + \frac{1}{\tilde{h}^2} \tilde{\nabla} \tilde{h} \cdot \tilde{\nabla} \tilde{\psi} + \frac{E_R^{1/2}}{\tilde{h}^2} \tilde{\mathcal{J}}(\tilde{h}, \tilde{\psi}), \tag{14}$$

where $\tilde{q} = (\tilde{\zeta} + 1)/\tilde{h}$ is the potential vorticity, $\tilde{\psi}$ is a modified streamfunction defined in Appendix C, $E_R = \nu/(\Omega R^2) = (h_0/R)^2 E$ is the Ekman number based on the radius, and $\tilde{F}$ is a forcing term described below. $\tilde{\mathcal{J}}$ is the non-dimensional Jacobian operator in cylindrical coordinates

$$\mathcal{J}(a, b) = \frac{1}{\tilde{\rho}} \left( \frac{\partial a}{\partial \tilde{\rho}} \frac{\partial b}{\partial \varphi} - \frac{\partial b}{\partial \tilde{\rho}} \frac{\partial a}{\partial \varphi} \right). \tag{15}$$

Equations (13)-(14) are solved for all the non-axisymmetric modes on the polar domain with free-slip boundary conditions for the outer and inner boundaries. The free-slip boundary conditions are employed to avoid high computational costs due to the accumulation of strong vorticity near the boundaries, particularly for the lowest $E$ cases. We verified that a no-slip boundary condition on the outer boundary leads to qualitatively similar results. For the axisymmetric mode $(\tilde{U}_\rho, \tilde{U}_\phi)$, we directly solve for the zonal flow evolution equation obtained by performing a zonal average of the zonal component of the Navier-Stokes equations (Aubert et al., 2003):

$$\frac{\partial \tilde{U}_\phi}{\partial \tilde{t}} + \tilde{U}_\rho \frac{\partial \tilde{U}_\phi}{\partial \tilde{\rho}} - \frac{\tilde{U}_\phi \tilde{U}_\rho}{\tilde{\rho}} + \tilde{U}_\rho = -\left\langle \tilde{u}'_\rho \frac{\partial \tilde{u}'_\phi}{\partial \tilde{\rho}} - \frac{\tilde{u}'_\phi \tilde{u}'_\rho}{\tilde{\rho}} \right\rangle_\phi + \frac{E_R}{2} \left( \tilde{\nabla}^2 \tilde{U}_\phi - \frac{\tilde{U}_\phi}{\tilde{\rho}^2} \right), \tag{16}$$

$$\tilde{U}_\rho = \frac{E_R^{1/2}}{2\tilde{h}} \tilde{U}_\phi. \tag{17}$$

The derivation of the previous equations as well as the numerical methods employed to solve them can be found in Appendix C.

Note that the simulations are performed with the same fluid height profile as in the experiment, $\tilde{h}(\tilde{\rho}) = \tilde{h}_{\min} \exp((|\beta| R/f)\tilde{\rho})$, i.e. with the same topographic $\beta$-effect. However, we use the expression of the Ekman pumping over a flat surface, and thereby neglect the bottom boundary curvature. This is justified by the fact that the experimental parameters were carefully chosen such that the bottom topography is as small as possible in amplitude (resulting in a maximum height difference of 5.36 cm and a mean absolute slope of 22%). Furthermore, the greatest slope is located at the center of the bottom plate, which is not resolved numerically. If the curvature of the bottom plate were to be considered, the Ekman pumping term would contain a geometrical factor involving the local slope of the bottom topography (Greenspan, 1968; Gillet, Brito, Jault and Nataf, 2007).

For now, we have introduced the forcing as an additional source of vorticity (term $\tilde{F}$ in equation (13)). The goal is to reproduce the experimental forcing such that the QG numerical model can be used as a guide and complement the experimental exploration. In the experiment, because of the Coriolis effect, each inlet or outlet generates respectively a small cyclone or anticyclone right above it. This process can be modeled as a stationary source of vorticity in the form of positive or negative Gaussian sources of vorticity of radius $\ell_f$ distributed on a prescribed array. We thus define $N$ forcing points distributed over the numerical domain, and at each point, we place a Gaussian source of vorticity such that

$$\tilde{F}(\tilde{x}, \tilde{y}) = \tilde{F}_0 \sum_{i=1}^{N} (-1)^i \exp\left(-\left[\frac{\tilde{x} - \tilde{x}_i}{\tilde{\ell}_f}\right]^2 - \left[\frac{\tilde{y} - \tilde{y}_i}{\tilde{\ell}_f}\right]^2\right), \tag{18}$$

where $(\tilde{x}, \tilde{y})$ are non-dimensional Cartesian coordinates, the pairs $(\tilde{x}_i, \tilde{y}_i)$, $i \in [[1, N]]$ are the center of each forcing vortex, $\tilde{\ell}_f$ is their radius and $\tilde{F}_0$ is the forcing amplitude. These vorticity sources are distributed over a prescribed stationary array. Both polar and Cartesian forcing arrays were tested. For the polar array, we chose the number of forcing rings arbitrarily, then, the vorticity sources are distributed such that the distance between two rings and between two adjacent vortices is approximately the same (this condition cannot be rigorously verified because of the periodicity of the domain in the azimuthal direction). There are as many positive as negative sources on each ring such that the zonally-averaged forcing term is zero by construction (no direct acceleration of the zonal flow). For the Cartesian forcing, we define the array on the square which encloses the circular domain. The number of lines of the Cartesian array is chosen a priori and the sign of the vorticity sources alternates along lines and columns. The array is then cropped to keep only sources fully within the circular domain. In any case, there is the same number of positive or negative vorticity sources such that there is no net angular momentum introduced by our forcing. In terms of amplitude, in the simulations, we work with a uniform amplitude over the domain. Experimentally, a uniform forcing requires to force more strongly at higher radii because the fluid height increases, which is why being able to control each forcing ring independently is important.

The parameters of the simulations discussed in the present paper are summarized in Table 4. We discuss one simulation with a polar forcing (QG1a) designed to match the experiment, two complementary simulations at smaller Ekman (QG1b,c) and three simulations with a Cartesian forcing and a progressively reduced distance between the vorticity sources to decrease the forcing scale compared to the experiment (QG2,3,4). In the following, all the numerical results are shown in dimensional form using the experimental parameters to allow for a direct comparison with the experimental measurements.

**Table 4**
Parameters of the quasi-geostrophic simulations used to complement the experimental results. $N$ is the number of vorticity sources, $\tilde{\Delta}_f$ is the average spacing between vorticity sources, $\ell_f$ is their radius, $F_0$ is the forcing amplitude (equation (18)) and $E_R$ the Ekman number based on the radius. For the Cartesian cases, $N$ is the number of sources on the square in which the disk in inscribed, and the number in parenthesis gives the exact number of sources on the disk. The experimental parameters are given as a reference, $\Delta_f$ and $\ell_f$ have been estimated by measuring the vorticity above forcing injection and suction points at the very beginning of an experiment (see Figure 9 in Lemasquerier et al. (2021)).

| Label | Forcing pattern | $N$ | $\tilde{\Delta}_f$ ($\Delta_f$) | $\tilde{\ell}_f$ ($\ell_f$) | $\tilde{F}_0$ ($F_0$) | $E_R$ |
|-------|-----------------|-----|-------------------------------|---------------------------|----------------------|-------|
| Exp | Polar | 128 (6 rings) | $1.4 \times 10^{-1}$ (7 cm) | $5.6 \times 10^{-2}$ (2.8 cm) | $2 \times 10^{-3}$ (0.5 $s^{-2}$) | $5.3 \times 10^{-7}$ |
| QG1a | Polar | 540 (12 rings) | $6.9 \times 10^{-2}$ (3.4 cm) | $1.1 \times 10^{-2}$ (5.4 mm) | $5 \times 10^{-3}$ (1.2 $s^{-2}$) | $1.2 \times 10^{-7}$ |
| QG1b | Polar | 540 (12 rings) | $6.9 \times 10^{-2}$ (3.4 cm) | $1.1 \times 10^{-2}$ (5.4 mm) | $5 \times 10^{-3}$ (1.2 $s^{-2}$) | $6 \times 10^{-8}$ |
| QG1c | Polar | 540 (12 rings) | $6.9 \times 10^{-2}$ (3.4 cm) | $1.1 \times 10^{-2}$ (5.4 mm) | $5 \times 10^{-3}$ (1.2 $s^{-2}$) | $2 \times 10^{-8}$ |
| QG2 | Cartesian | $24^2$ (408) | $8.3 \times 10^{-2}$ (4.2 cm) | $1.1 \times 10^{-2}$ (5.4 mm) | $5 \times 10^{-3}$ (1.2 $s^{-2}$) | $1.2 \times 10^{-7}$ |
| QG3 | Cartesian | $48^2$ (1716) | $4.2 \times 10^{-2}$ (2.1 cm) | $1.1 \times 10^{-2}$ (5.4 mm) | $5 \times 10^{-3}$ (1.2 $s^{-2}$) | $6.1 \times 10^{-8}$ |
| QG4 | Cartesian | $72^2$ (4404) | $2.8 \times 10^{-2}$ (1.4 cm) | $1.1 \times 10^{-2}$ (5.4 mm) | $5 \times 10^{-3}$ (1.2 $s^{-2}$) | $4.1 \times 10^{-8}$ |

## 3. Qualitative observations

### 3.1. Experimental flows

After reaching solid-body rotation, the forcing is turned on to reach uniform rms velocity fluctuations over the domain. We work with forcing amplitudes above the threshold of the transition identified in Lemasquerier et al. (2021). In this regime, zonal jets progressively emerge from the forced turbulent flow, with a transient lasting about 1000 rotation periods (800s), as seen on the space-time diagram of Fig.12(a). During this transient, the jets, which initially emerge at the scale of the forcing rings, drift and merge to self-organize at a larger scale. A statistically-steady state is then achieved where three to one prograde jets are observed depending on the forcing amplitude. Fig.3*(a,c)* show two instantaneous velocity fields as measured by PIV once the statistically steady state is achieved. Both experiments are in Regime II but correspond to different forcing amplitude, leading to $Re = 10.8 \times 10^3$ and $19.7 \times 10^3$ respectively (experiments J and A of Table 3). Qualitatively, our experiments performed at higher Reynolds number lead to broader and intensified zonal jets, as can be seen in Fig.3*(b,d)* which shows the zonal flow profile for the same experiments. Since the prograde jets width increase with the forcing amplitude, for our most extreme experiments, only one prograde jet can fit in the experimental domain, whereas two to three prograde jets can be obtained for the smallest $Re$. As the forcing is increased, we also notice that the prograde jets become less "wavy". Our hypothesis is that for moderate $Re$, the flow is close to the transition threshold and thus quasi-resonant, as described in Lemasquerier et al. (2021). Farther from the transition threshold, the resonant Rossby waves are less prone to develop and interact with the jet.

In Lemasquerier et al. (2021), we show that in a classical quasi-geostrophic model of the experiment, the zonal flow evolution equation (16) reduces to

$$\frac{\partial U_\phi}{\partial t} = -\frac{1}{\rho^2} \frac{\partial \langle \rho^2 u'_\rho u'_\phi \rangle_\phi}{\partial \rho} + \mathcal{D}. \tag{19}$$

Here, $\mathcal{D}$ represents both bulk and parietal viscous dissipation. The first term on the right-hand-side is the Reynolds stresses divergence, and is responsible for accelerating the zonal flow. In Lemasquerier et al. (2021), we show that at the very beginning of an experiment, the zonal jets emerge following convergence of prograde momentum at the location of the forcing rings because of the radiation of Rossby waves (see Figure 9*(d)* in Lemasquerier et al. (2021)). Once in a fully turbulent, quasi-steady state, we loose this very regular pattern seen in the transient and the associated velocity correlations. However, the zonal jets remain eddy-driven, as illustrated on Figure 4, where we represent the time-averaged radial profile of the Reynolds stresses, for experiment J and the reference simulation QG1a. It is clear that the Reynolds stresses decrease in a prograde jet (prograde momentum convergence), and increase in retrograde flows (prograde momentum divergence).

## 3.2. Independence on the forcing scale

The QG numerical simulations can be used to verify that the zonal flow in Regime II is not sensitive to the detail of the forcing pattern in terms of length scale and distribution (polar versus Cartesian).

The first step was to find a numerical setup for which the obtained solution reproduces well the experimental observations. With this in mind, we performed simulations with a polar forcing pattern. The first important result is that in our QG numerical simulations, we are able to retrieve the transition between the two regimes identified experimentally in Lemasquerier et al. (2021), with a transition between locally and globally forced jets (not shown). As this transition is not the focus of the present paper, we focus only on simulations in Regime II. By slightly tuning the forcing scale and amplitude, we define a reference simulation which reproduces best the experimental flow, labelled QG1a in Table 4. The corresponding space-time diagram is shown in Appendix A, Fig.12*(c)*, and shows the emergence of zonal jets with properties very close to the experimental ones. Fig.5 shows the vorticity and azimuthal velocity fields for experiment J and four QG simulations, once in the statistically steady state. Comparing panels *(a)* and *(b)* of Fig.5 shows that the simulation QG1a exhibits three prograde jets of similar amplitude and width compared to the experiment. This reference simulation shows that for the QG simulations to best reproduce the experimental results, we should use forcing patterns with a typical scale smaller than the experimental pattern (540 inlets and outlets arranged on 12 forcing rings in the present case). This is not surprising given that the QG model imposes a purely barotropic and stationary forcing, whereas in the experiment, three-dimensional effects affect the forcing injections and suctions which are not perfect gaussian sources of vorticity aligned along the rotation axis, and which are very likely to generate smaller and more fluctuating structures.

Fig.5*(c)* shows the results of simulation QG2 performed with a Cartesian forcing pattern, of approximately the same length scale as in the polar case discussed above, QG1a. The resulting statistically steady state is similar to what is obtained with a polar array. This demonstrates that the polar distribution of the forcing does not significantly influence the late-time jet profile. We note however that whereas the jets do not drift in the experiments and the polar simulation QG1a, the cartesian forcing seems to allow for a slow radial drift of the prograde jets (Fig.12*(e)*). This is probably because the forcing cannot be strictly equal on both sides of each jet given its Cartesian distribution. This might also explain the drift reported in Cabanes et al. (2017) where a Cartesian forcing pattern was used in the experiment. We wish to underline that in our cartesian forcing simulations, the jets drift even if the $\beta$-effect is uniform. Hence, the

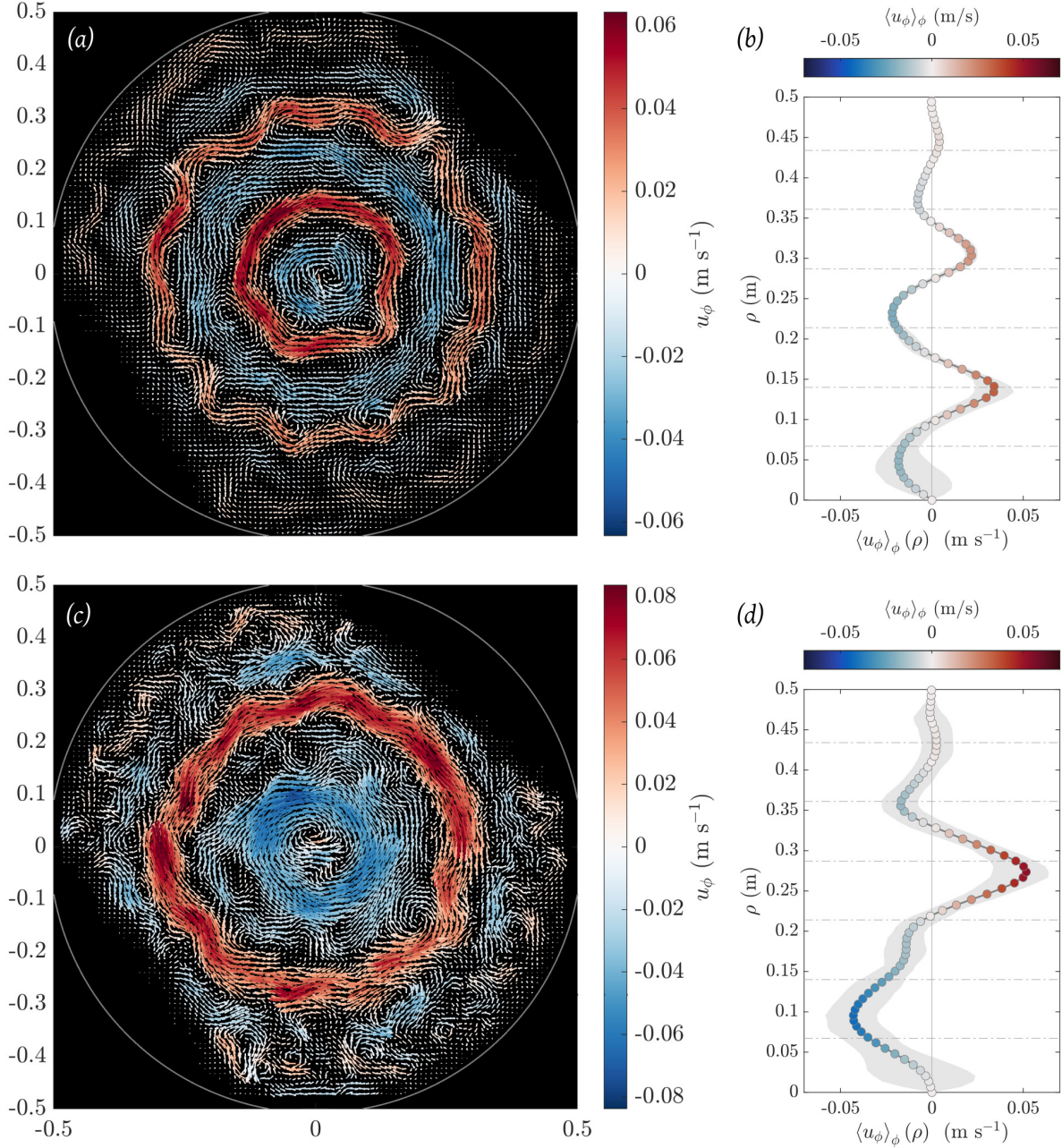

**Figure 3:** *(a,c)* Instantaneous velocity fields measured from PIV in the statistically steady state of *(a)* experiment J and *(c)* experiment A of Table 3. The vectors magnitude scale is the same for the two plots. *(b,d)* Time-averaged zonally-averaged zonal flow in *(b)* experiment J and *(d)* experiment A of Table 3. The shaded area represents the envelope of all the instantaneous profiles. Dashed lines represent the radii of the forcing rings.

drift mechanism should be captured with models in the $\beta$-plane approximation (see e.g., Cope (2021)). Second, while useful experimentally, it appears that the polar forcing pattern artificially locks the jets at a fixed radial position. Note that we performed experiments where the power of each forcing ring fluctuates randomly around a prescribed mean, and experiments where the forcing alternates between two groups of three rings, but we never observed a long-term drift.

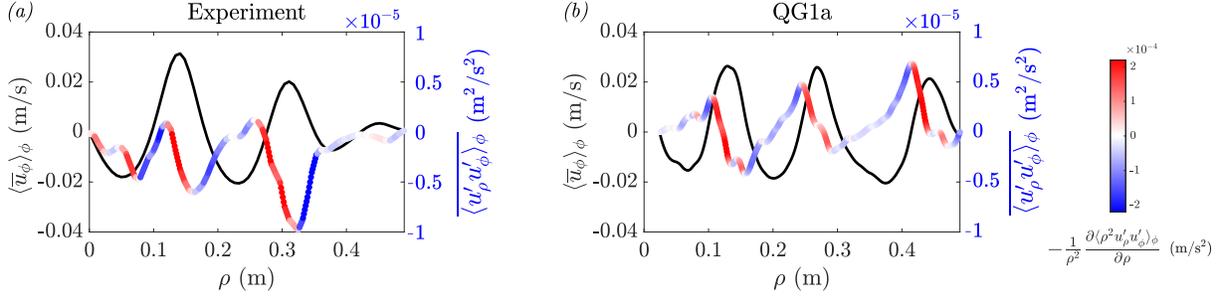

**Figure 4:** Comparison between the zonal flow (black line) and the Reynolds stresses or eddy momentum fluxes (colored line) in the experiment J *(a)* and the simulation QG1a *(b)*. The left *y*-axis corresponds to the zonal flow velocity, and the right *y*-axis is for the Reynolds stresses. The colour of the line represents the divergence of the Reynolds stresses. When it is positive (red), it indicates a convergence of prograde momentum, and a divergence when it is negative (blue).

Finally, Fig.5*(d,e)* show the result of simulations performed with a Cartesian pattern of smaller forcing scale (QG3 and QG4). We recall that for these simulations, we decrease the forcing scale by dividing the distance between vorticity sources by a factor 2 (see Table 4). The decrease in the forcing scale is evident on the vorticity maps which exhibit finer structures, however, qualitatively, the late-time jets amplitude and width do not vary significantly. Note that comparing the number and position of the prograde jets is not robust, given that the system is multistable: as illustrated in Fig.6, the number and position of the jets can vary for different realisations of the exact same simulation. This multistability is also observed in the experiment. Hence, we argue that the fact that simulation QG4 loses the most central prograde jet is not significant. More generic, idealized simulations and systematics should be performed to make this result more robust. Notably, Scott and Dritschel (2019) demonstrated in the framework of potential vorticity mixing that the late-time resulting zonal jets profile is the same whether the forcing is performed at large scale or at small scale, the important parameter being the zonostrophy index $R_\beta$ only. This study could be extended to our geometry, but this is beyond the scope of the present work. Here, our point is that modifying the experimental setup to perform a forcing at smaller scale does not seem to be extremely valuable, since the jets dynamics appears to be unaffected by changes to the forcing scale.

## 4. Spectral analysis

### 4.1. Spectral properties and signature of the zonostrophic regime

We now turn to a spectral analysis of the flow observed experimentally and numerically. The goal of this section is to assess whether Regime II is consistent with the predictions in the regime of zonostrophic turbulence (Galperin et al., 2010), and hence relevant to the regimes of turbulent zonal jets observed on Jupiter (Choi and Showman, 2011; Galperin et al., 2014b). This is not evident a priori given that zonostrophic turbulence was described from simulations of two-dimensional turbulence on the sphere. Whether such regime can be achieved in a fully three-dimensional flow – yet constrained by rotation – still remains to be addressed.

#### 4.1.1. Zonal and residual kinetic energy spectra

In the zonostrophic turbulence regime, the turbulent flow shows profound anisotropy, and the kinetic energy spectra computed from the axisymmetric component of the flow $E_z$ (hereafter, zonal spectrum) and from the residual non-axisymmetric component $E_r$ (residual spectrum) follow different scalings similar to those in $\beta$-plane turbulence (Huang, Galperin and Sukoriansky, 2000; Sukoriansky et al., 2002; Galperin et al., 2006; Sukoriansky et al., 2007). Assuming isotropy for intermediate wave number modes (scales smaller than the transitional scale but larger than the forcing scale), the residual spectrum is expected to be compatible with the inverse cascade branch of the Kolmogorov-Batchelor-Kraichnan theory of two-dimensional turbulence:

$$E_r^{\text{theo}}(k) \sim C_K \epsilon^{2/3} k^{-5/3}, \tag{20}$$

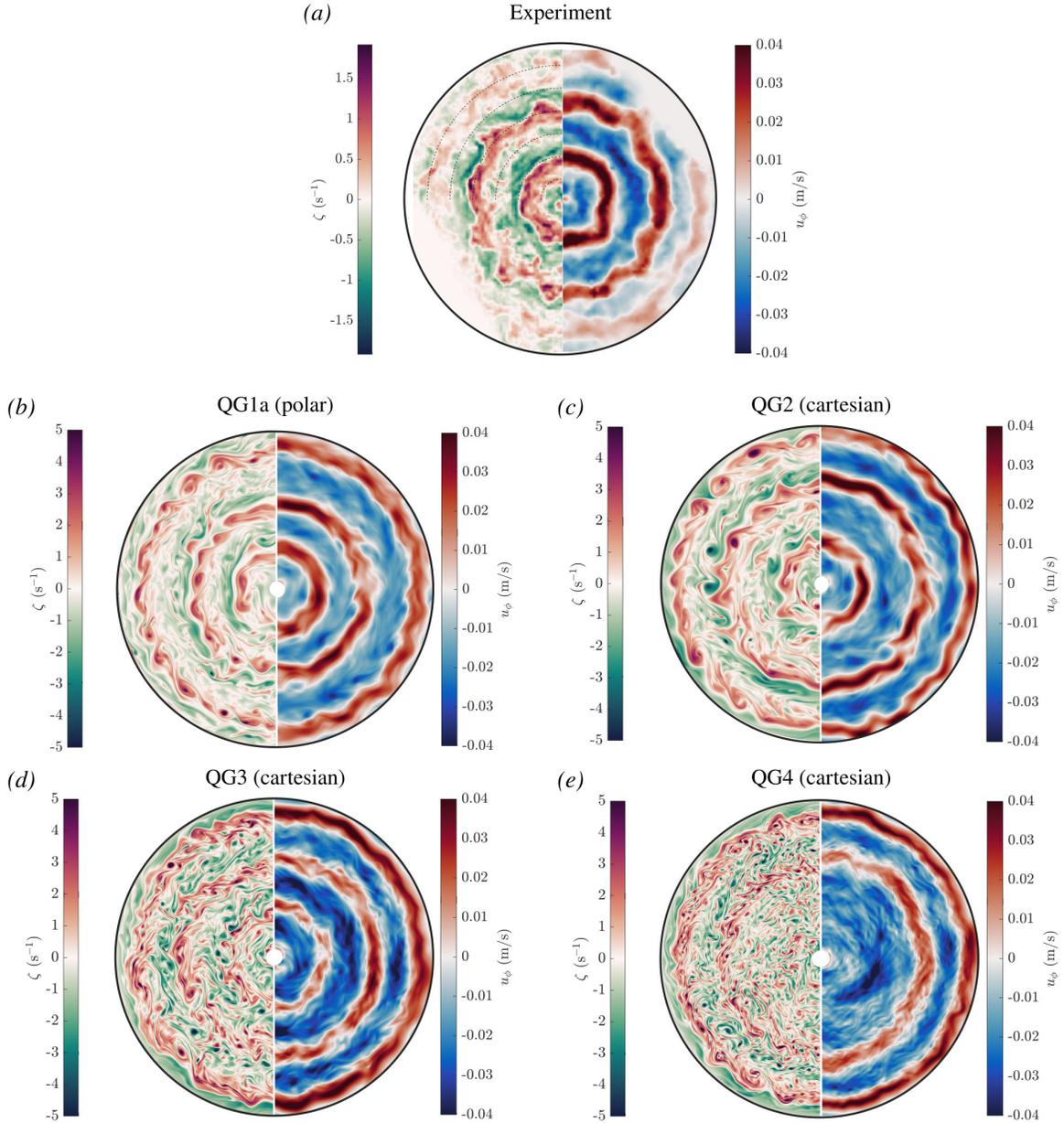

**Figure 5:** Instantaneous maps of vorticity $\zeta$ (left side of each panel) and azimuthal velocity $u_\phi$ (right side of each panel). *(a)* Experiment J. *(b)* References simulation QG1a with a polar forcing. *(c,d,e)* Cartesian forcing simulations (QG2, QG3 and QG4) with a progressively reduced forcing scale.

where $k$ is the total wavenumber, and $C_K \approx 6$ is the universal Kolmogorov-Kraichnan constant (Boffetta and Ecke, 2012). On the contrary, the zonal spectrum follows a steeper slope:

$$E_z^{\text{theo}}(k_r) \sim C_Z \beta^2 k_r^{-5}, \tag{21}$$

with $k_r$ the wavenumber in the direction orthogonal to the zonal flow, i.e. the radial wavenumber in our case. $C_Z$ is a constant of order unity (Chekhlov, Orszag, Sukoriansky, Galperin and Staroselsky, 1996; Huang et al., 2000), whose value was shown to lie around 0.5 by numerical simulations on the sphere (Sukoriansky et al., 2002), around 2 for Jupiter (Galperin et al., 2014b) and from 1.7 to 3.7 in the experiments of Cabanes et al. (2017). The theory of zonostrophic turbulence shows that the kinetic energy spectra can provide a useful insight on the exchange of energy between scales as well as a diagnostic tool to determine to what extent a given flow is in the regime of zonostrophic turbulence. In this section, we compare the predictions from this theory with our experimental measurements, complemented by QG simulations.

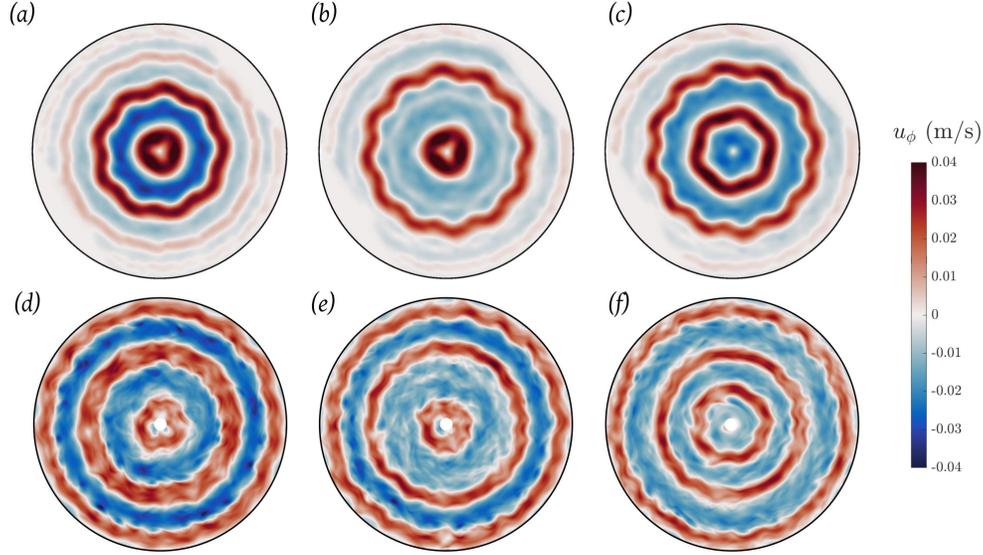

**Figure 6:** Maps of azimuthal velocity once in steady state, illustrating the multistability. *(a-c)* Experiments: three realizations with the same forcing as Exp.J. *(d-f)* QG simulations: three realizations with the parameters of simulation QG1a. The initial condition of the simulations is a fluid at rest, plus a random noise of non-dimensional amplitude $1 \times 10^{-3}$ added to the vorticity field. For each realization, the seed of the initial noise was changed.

Given the rotational symmetry of the experimental 2D velocity fields obtained from PIV, we perform a Bessel-Fourier decomposition relevant to the polar system of coordinates (Wang, Ronneberger and Burkhardt, 2008), for which details can be found in the Appendix D. Owing to the periodicity in $\phi$, the finite domain in radius, $\rho \leq R$, and the zero-value boundary condition at $\rho = R$, the relevant basis functions $\Psi_{nm}$ to decompose our fields are separable in polar coordinates and consist in a complex exponential for the angular part, and normalized Bessel functions for the radial part:

$$\Psi_{nm}(\rho, \phi) = \frac{1}{\sqrt{2\pi N_{nm}}} J_m(k_{nm}\rho) e^{im\phi}, \tag{22}$$

$$\text{with} \quad N_{nm} = \frac{R^2}{2} J_{m+1}^2(\alpha_{nm}). \tag{23}$$

$J_m$ is the Bessel function of the first kind of order $m$, $\alpha_{nm}$ is the $n^{\text{th}}$ positive zero of $J_m$, and the radial wavenumber $k_{nm} = \alpha_{nm}/R$ takes discrete values since the domain is bounded. For each $\Psi_{nm}$, $m$ is the number of periods in the angular direction, and $n-1$ corresponds to the number of zero crossings in the radial direction. The value of $k_{nm}$ is thus an indication of the scale of the basic patterns, similarly to the normal Fourier transform. Any function, including our velocity fields $\boldsymbol{u} = (u_\rho, u_\phi)$ can be decomposed on this basis such that

$$\boldsymbol{u}(\rho, \phi) = \sum_{n=1}^{\infty} \sum_{m=-\infty}^{+\infty} \hat{\hat{\boldsymbol{u}}}_{nm} \Psi_{nm}(\rho, \phi), \tag{24}$$

where the double hat indicates the double transform. The Bessel-Fourier transform coefficients are

$$\hat{\hat{\boldsymbol{u}}}_{nm} = \int_0^R \int_0^{2\pi} \boldsymbol{u}(\rho, \phi) \Psi_{nm}^*(\rho, \phi) \rho d\rho d\phi, \tag{25}$$

where the star denotes the complex conjugate. For our experimental results, the PIV fields are obtained on a regular Cartesian grid. To evaluate the coefficients $\hat{u}_{nm} = (\hat{u}_{\rho,nm}, \hat{u}_{\phi,nm})$, we perform a spline interpolation of the velocity field on a polar grid where the radial and azimuthal parts of the transform are separable. Since the velocity field is discrete, the coefficients (25) are computed using fast Fourier transform and discrete Hankel transform algorithms. The angular part is done using the Matlab *fft* function, and the radial part is performed using the Matlab algorithm provided by Guizar-Sicairos and Gutiérrez-Vega (2004). Using the Parseval relation which arises from the orthogonality of the basis functions, kinetic energy spectra can be computed directly in the spectral space, and are proportional to $\hat{u}_{nm}\hat{u}_{nm}^*$ (see Appendix D). We denote $E_{nm}$ the kinetic energy contained at the wavenumber $k_{nm}$. We distinguish the kinetic energy contained in the zonal mode characterized by $m = 0$ and denoted $E_z$, and the residual kinetic energy spectra $E_r$, which is the sum of the contribution of all the non-zonal modes, $m \neq 0$.

Fig.7(a,b) show the zonal and residual kinetic energy spectra for experiments M and B of Table 3. The spectra are computed once the statistically steady state is achieved and time-averaged over 50 statistically-independent spectra spanning 600 rotation times. Exp. B has a Reynolds number about twice higher compared to Exp. M (*Re* ~15,900 versus 7310). The energy spectra provide different information depending whether we are looking at scales smaller or larger than the forcing scale. Experimentally, the forcing scale should lie around the mean distance between two injections on the bottom plate, i.e. $\Delta_f \approx 7$ cm ($k_f \sim 90$ rad m$^{-1}$). At scales smaller than the forcing scale (shaded area, $k > k_f$), the residual spectra exhibit a slope close to $k^{-4}$. This slope is steeper than the $-3$ slope expected from the direct cascade of enstrophy, but this behavior is commonly observed in both numerical simulations and experiments of 2D turbulence, and may be due either to friction effects, or to the fact that we do not use any logarithmic correction (see Boffetta and Ecke, 2012, for a review). At scales larger than the forcing scale ($k < k_f$). We observe that the residual energy spectra are compatible with a $-5/3$ slope consistent with the presence of an inverse cascade of energy. This slope stops as soon as $k_\beta$ is reached, meaning that the isotropic inverse cascade does not continue past the anisotropisation threshold. The dark green and blue lines in Fig.7(a,b) show the zonal energy spectra. During the transient (thin dark blue line in panel *(b)*), the zonal energy first peaks close to the forcing scale, meaning that the zonal flow is first locally forced. Once in the statistically steady state and for both experiments M and B, the zonal spectra peak at a larger scale meaning that there is a progressive energy transfer towards scales larger than the forcing scale. As predicted in the zonostrophic regime, the spectrum follows a steep $k^{-5}$ slope. This means that the zonal flow profile is extremely smooth with localised sharp features leading to a rapidly decaying spectral slope. We note that the $k^{-5}$ slope seems to continue below the forcing scale. Even if this was also observed in previous studies (Zhang and Afanasyev, 2014; Cabanes et al., 2017; Cabanes, Favier and Le Bars, 2018), we argue that it is not a robust feature given that it corresponds to scales smaller than the energy injection scale. For experiments M and B, the zonal energy peaks at $k \sim 30$ rad m$^{-1}$ and 24 rad m$^{-1}$ respectively, corresponding to wavelengths of 21 cm and 26 cm which represent the large radial wavelength of the zonal jets. At the largest scales, the zonal $m = 0$ mode contains more energy than all the residual modes. More precisely, the time-averaged total kinetic energy in the flow is of $7.94 \times 10^{-5}$m$^2$ s$^{-2}$ in Exp. M and the zonal mode contains 58% of it, whereas for Exp. B, the total kinetic energy is of $3.73 \times 10^{-4}$m$^2$ s$^{-2}$ and the zonal mode contains 68% of it. This is significantly larger than our experiments in Regime I, where the zonal flow contains at most ~ 20% of the total kinetic energy (not shown, see Lemasquerier et al. (2021) for more details).

### 4.1.2. Rate of upscale energy transfer

Previously, we estimated the upscale energy transfer of the inverse cascade, $\epsilon^E$, by assuming that dissipation consists in Ekman friction only (equation (7)). The amplitude of the theoretical residual spectra can be used to actually measure $\epsilon$ (equation (20)) which is otherwise difficult to do experimentally. To do so, we take the commonly accepted value $C_K \approx 6$ for the Kolmogorov-Kraichnan constant (Boffetta and Ecke, 2012) and deduce $\epsilon$ from the fit of the $-5/3$ slope on the spectra. The deduced values, denoted $\epsilon^S$, are reported in Table 5, and have the correct order of magnitude compared to the estimated $\epsilon^E$ (Table 3). The precise comparison between the two, shown in Fig.8(a), shows a small deviation from a pure proportionality relationship. For the zonal spectra, since $\beta$ is known, we report in Table 5 the values of $C_Z$ corresponding to our fits (equation (21)). $C_Z$ is supposedly universal in the zonostrophic regime. According to the numerical simulations reported in Sukoriansky et al. (2002); Galperin et al. (2006); Sukoriansky et al. (2007), it should lie around 0.5. Our experimental measurements show that $C_Z \sim 0.1 - 0.3$ for all of our experiments despite very different upscale energy transfer rate, which confirms the idea of a universal constant. Its value is smaller for our weakest experiments (L,M,N), probably because these experiments are only slightly super-critical with respect to the transition from Regime I to II described in Lemasquerier et al. (2021). We also note that our value for $C_Z$ is smaller than that of Cabanes et al. (2017) who found $C_Z \sim 1.7 - 3.7$. This difference may come from the fact that the experiments of Cabanes et al. (2017) are not in the $\beta$-plane framework since $\beta$ is strongly varying with radius. The $\beta$ parameter to use in equation (21) is thus uncertain, whereas in our case $\beta$ is uniform and uniquely defined. In addition, the forcing scale in Cabanes et al. (2017) is twice larger than the present one, and close to the transitional scale as we discuss later. Consistently, Galperin et al. (2006) report that in the case of a too small scale separation between the forcing scale and the transitional scale, the spectra exhibit non-universal behaviors.

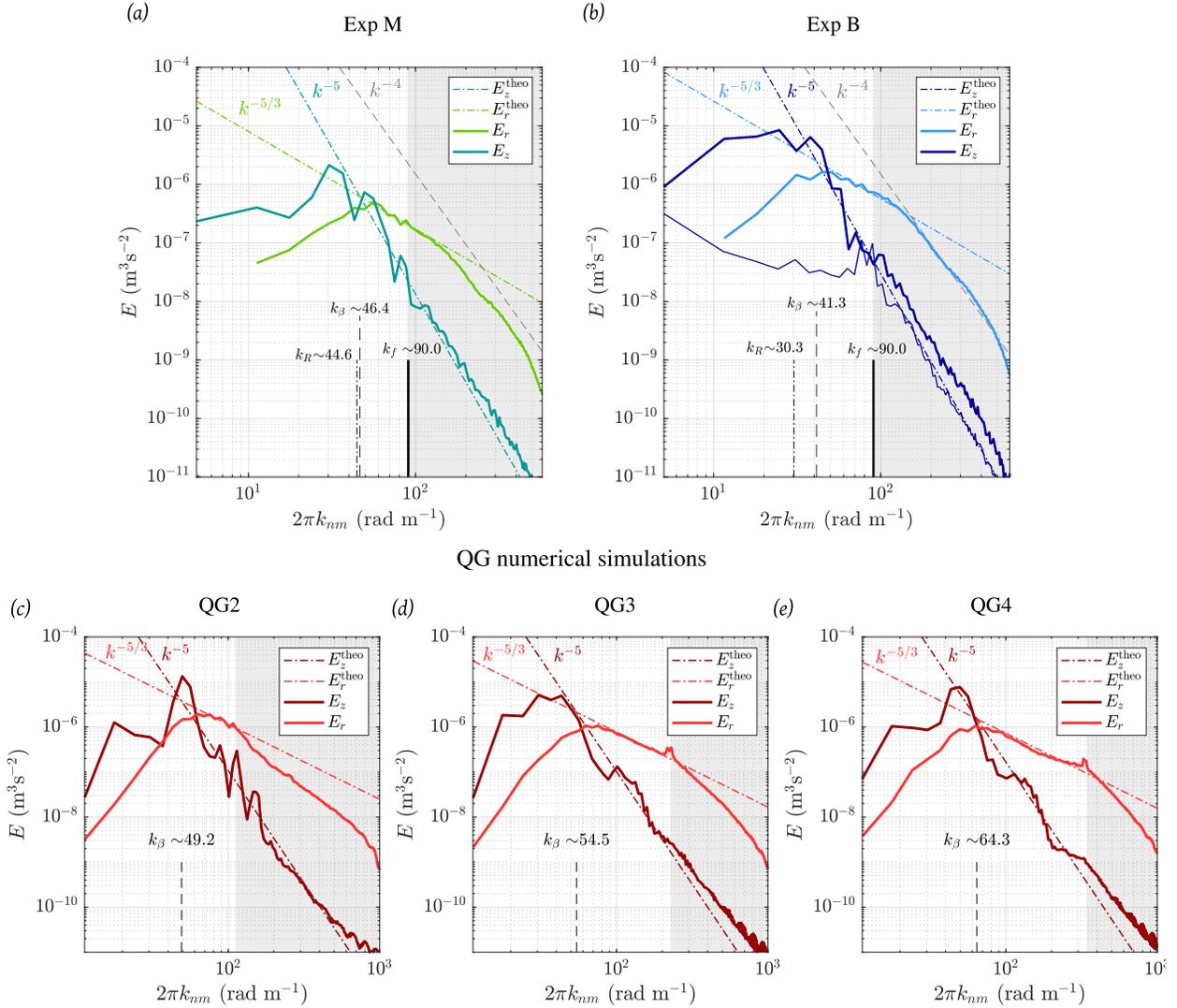

**Figure 7:** Kinetic energy spectra for two different experiments and three QG numerical simulations. We separate the zonal ($E_z$) and residual ($E_r$) contributions. The dashed-dotted lines correspond to the theoretical predictions given by equations (20) and (21). The shaded area correspond to scales smaller than the forcing scale, $k > k_f$. *(a)* Experiment M of Table 3. *(b)* Experiment B of Table 3. The thin dark blue lines correspond to $E_z$ computed during the transient of the experiment. For $E_r^{\mathbf{theo}}$, we take $C_K = 6$ and find $\epsilon$ that best fits the experimental spectra. For $E_z^{\mathbf{theo}}$, since $\beta$ is known, we find $C_Z$ that best fits the data. *(c–e)* Spectra derived from QG numerical simulations with decreasing forcing scale compared to the experiment.

To complement the experiments, we performed the same spectral analysis on the QG simulations. The results are reported in Table 5. The parameters measured on the reference simulation, QG1a, are quantitatively consistent with the spectral parameters measured from the experimental results, for both $\epsilon^S$ and $C_Z$.

---

**Table 5**
Parameters measured from the spectral analysis. We use $C_K = 6$ and deduce $\epsilon^S$ from equation (20) using a fit of the residual spectra for wavenumbers between the forcing wavenumber and the peak energy wavenumber. The $-5/3$ slope of the fit is imposed, and the prefactor is determined using Matlab's *fit* function[a] with least squares fit type. The confidence interval corresponds to the 95% confidence bounds. The constant $C_Z$ is determined by a fit on the $-5$ slope of the zonal spectra, since $\beta$ is known (equation (21)). The transitional wavenumber based on the spectral slopes intersection is denoted $k_\beta^S$ whereas the transitional wavenumber based on $\epsilon$ only is denoted $k_\beta^\epsilon$ (equation (26)). $k_R$ is the Rhines wavenumber (equation (2)), $R_\beta^S = k_\beta^S/k_R$ and $R_\beta^\epsilon = k_\beta^\epsilon/k_R$. $L_T$ is the Thorpe scale measured experimentally and discussed in section 5.

| Label | $\epsilon^S$ ($\times 10^{-7}\,\mathrm{m^2\,s^{-3}}$) | $C_Z$ | $k_\beta^S$ ($\mathrm{rad\,m^{-1}}$) | $k_\beta^\epsilon$ ($\mathrm{rad\,m^{-1}}$) | $k_R$ ($\mathrm{rad\,m^{-1}}$) | $R_\beta^S$ | $R_\beta^\epsilon$ | $L_T$ (cm) |
|---|---|---|---|---|---|---|---|---|
| A | $70.8 \in [61.3, 81.7]$ | 0.32 | 46.3 | 112.2 | 27.2 | 1.72 | 4.16 | 2.61 |
| B | $30.3 \in [26.5, 34.7]$ | 0.27 | 52.6 | 132.9 | 30.5 | 1.73 | 4.39 | 2.11 |
| C | $55.6 \in [43.8, 70.5]$ | 0.13 | 43.9 | 138.3 | 32.1 | 1.38 | 4.35 | 2.11 |
| D | $96.4 \in [78.9, 117]$ | 0.86 | 36.7 | 65.7 | 18.2 | 2.06 | 3.69 | 3.51 |
| E | $47.0 \in [42.2, 52.4]$ | 0.27 | 48.3 | 121.7 | 29.1 | 1.67 | 4.22 | 2.23 |
| F | $66.3 \in [58.2, 75.5]$ | 0.27 | 44.6 | 113.7 | 27.6 | 1.63 | 4.16 | 2.72 |
| G | $49.0 \in [41.6, 57.7]$ | 0.25 | 46.5 | 120.7 | 27.6 | 1.69 | 4.39 | 2.60 |
| H | $7.69 \in [6.87, 8.61]$ | 0.13 | 56.5 | 174.8 | 39.3 | 1.44 | 4.47 | 1.99 |
| I | $8.73 \in [5.78, 13.2]$ | 0.21 | 62.7 | 170.5 | 36.9 | 1.70 | 4.63 | 1.57 |
| J | $8.37 \in [5.66, 12.4]$ | 0.18 | 60.4 | 171.9 | 36.6 | 1.65 | 4.71 | 1.80 |
| K | $5.09 \in [4.24, 6.13]$ | 0.38 | 51.8 | 118.4 | 26.7 | 1.96 | 4.49 | 2.56 |
| L | $4.44 \in [4.03, 4.90]$ | 0.11 | 59.3 | 195.1 | 46.1 | 1.29 | 4.24 | 1.38 |
| M | $4.78 \in [4.17, 5.47]$ | 0.10 | 56.2 | 192.3 | 45.0 | 1.26 | 4.32 | 1.35 |
| N | $1.70 \in [1.12, 5.83]$ | 0.06 | 59.6 | 236.4 | 52.2 | 1.15 | 4.54 | 0.98 |
| QG1a | $25.4 \in [20.7, 31.1]$ | 0.25 | 53.3 | 137.7 | 30.6 | 1.61 | 3.55 | 1.89 |
| QG1b | $49.8 \in [36.1, 68.8]$ | 0.57 | 59.6 | 120.4 | 32.3 | 1.69 | 3.78 | 2.53 |
| QG1c | $20.0 \in [16.2, 24.9]$ | 0.37 | 62.5 | 144.4 | 31.80 | 2.02 | 3.91 | 1.99 |
| QG2 | $84.1 \in [69.3, 102]$ | 0.43 | 49.2 | 108.4 | 33.6 | 1.58 | 4.09 | 2.38 |
| QG3 | $46.2 \in [41.8, 51.1]$ | 0.41 | 54.5 | 122.2 | 28.6 | 2.08 | 4.21 | 2.94 |
| QG4 | $42.3 \in [38.0, 47.1]$ | 0.67 | 64.3 | 124.4 | 29.07 | 2.15 | 4.97 | 2.74 |

[a]https://mathworks.com/help/curvefit/fit.html

### 4.1.3. Zonostrophy index

The intersection of the theoretical $-5/3$ (equation (20)) and $-5$ (equation (21)) slopes defines a spectral transitional wavenumber $k_\beta^S$, which can be expressed as

$$k_\beta^S = \left(\frac{C_Z}{C_K}\right)^{3/10} \underbrace{\left(\frac{\beta^3}{\epsilon^S}\right)^{1/5}}_{k_\beta^\epsilon}. \tag{26}$$

This scale thus defines the threshold of spectral anisotropy. In our case, since $C_K = 6$ and $C_Z \sim 0.1 - 0.3$, the prefactor $(C_Z/C_K)^{3/10}$ is thus of about $0.29 - 0.41$. The spectral transitional wavenumber, $k_\beta^S$ is thus less than half that based on the values of $\epsilon^S$ and $\beta$ only, denoted $k_\beta^\epsilon$. Both $k_\beta^S$ and $k_\beta^\epsilon$ are reported in Table 5 for our selected experiments. We additionally report the values of the Rhines wavenumber $k_R$, based on the total rms velocity. The Rhines and transitional scales are also indicated on the experimental spectra, Fig.7(a,b). Finally, the last two columns of Table 5 report the zonostrophy index with or without using the prefactor on the transitional wavenumber: $R_\beta^S = k_\beta^S/k_R$ and $R_\beta^\epsilon = k_\beta^\epsilon/k_R$. The spectral zonostrophy index of our experiments lies between 1.15 and 2.06, whereas without the prefactor on $k_\beta$, we obtain values between 3.69 and 4.71. The QG simulations again support this result, with $R_\beta^S \sim 1.61$ and $R_\beta^\epsilon \sim 3.55$ for the reference simulation QG1a.

The moderate values of $R_\beta^S$ deduced from our spectral analysis question the extent to which our experiments are in the zonostrophic regime. Using 2D barotropic simulations on the sphere, Galperin et al. (2006); Sukoriansky et al. (2007) found $C_Z \sim 0.5$ and $C_K \sim 6$ (prefactor $(C_Z/C_K)^{3/10} \sim 0.5$), and a lower bound for the zonostrophic regime $R_\beta^S \gtrsim 2.5$. With values below 2.5, we would then be in the transitional regime between the dissipation-dominated regime and the zonostrophic turbulence. But since we measure $C_Z \sim 0.1 - 0.3$, the prefactor is reduced, and the threshold should be rescaled as $R_\beta^S \gtrsim 1.8$. In addition, the 2.5 threshold in $R_\beta$ is only indicative (the boundary between the regimes is not a strict proportionality relationship, and implies an offset, see Fig.13.2 in (Galperin et al., 2019)). We also recall that the boundaries between zonostrophic and friction-dominated regimes have been obtained in the very specific case of a purely two-dimensional flow on the sphere, which significantly differs from our setup. That being said, in Galperin et al. (2006); Sukoriansky et al. (2007), the authors in fact use the spectra as a diagnostic tool to determine the flow regime. In other words, the simulations considered in the zonostrophic regime are those where the predicted spectra are recovered, with a *universal* behavior and a $C_Z \sim 0.5$ constant. Consistently, we argue that one should rather look at the spectra to determine in which regime we stand, and not the absolute value of the zonostrophy index alone. The important differences between our 3D experiments and their 2D simulations on the sphere is very likely to induce significant differences in the threshold of the zonostrophic regime. The fact that we recover a $-5$ steep slope for the zonal spectrum, and that the zonal flow contains more energy than the fluctuations at large scales is somewhat more robust and physically meaningful that the absolute value of $R_\beta$ alone. The zonostrophy index is nevertheless useful to compare the degree of zonostrophy of different experiments, as soon as it is computed the same way for each experiment. This is what was done to plot Fig.1, where we compare various experiments on zonal jets.

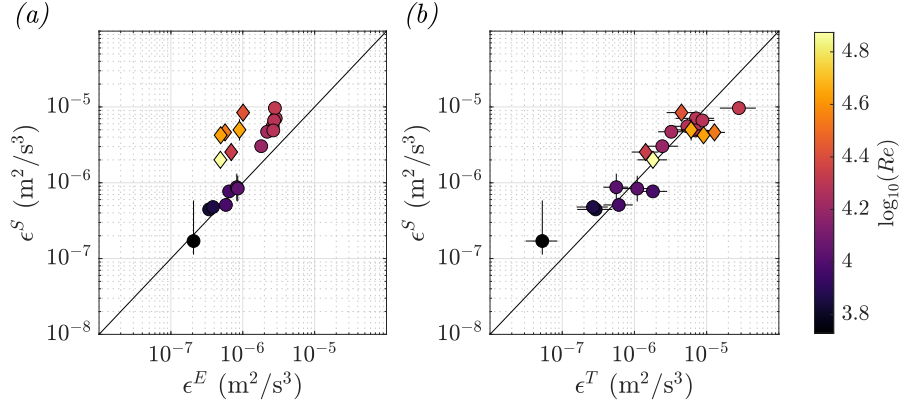

**Figure 8:** Comparison between estimates of the upscale energy transfer rates. Circles: experiments. Diamonds: QG simulations. *(a)* Upscale energy transfer rate measured on the spectra, $\epsilon^S$, versus the one estimated assuming pure dissipation by Ekman friction, $\epsilon^E$ (equation (7)). *(b)* $\epsilon^S$ versus the transfer rate deduced from the Thorpe scale, $\epsilon^T$, assuming that $L_\beta^T = L_T/0.47$ (Fig.11(b)) and $\epsilon^T = (L_\beta^T/2\pi)^5\beta^3$. Vertical error bars account for the uncertainty in $\epsilon^S$ due to the uncertainty in the slope measured on the spectra (see Table 5). Horizontal error bars account for the standard deviation when measuring the Thorpe scale.

### 4.2. Influence of the forcing scale

An important point that we want to discuss is the influence of the forcing scale. Galperin et al. (2006) underline that the zonostrophic regime is not defined by a single inequality on $R_\beta$, but rather a series of inequalities expressed as: $30/R \leq 8k_E \leq 2k_\beta \leq k_f$, in their specific 2D simulations on the sphere, where $R$ is the radius of the sphere, and we recall that $k_f$ is the forcing wavenumber. The last inequality requires the forcing to act at a scale smaller than half the scale at which the eddies start to feel the $\beta$-effect. Physically, this constraint can be seen as the need for a significant Kolmogorov-Kraichnan inertial range to exist such that an isotropic inverse cascade is able to develop between $k_f$ and $k_\beta$. Practically, when the forcing scale is too large, Galperin et al. (2006) report that non-universal behaviors are observed on the spectra, without giving much additional details. In environmental flows (oceans and planetary atmospheres), typically, the forcing acts at a scale smaller than the scale of turbulence anisotropisation $L_\beta$ by a factor 2 to 3 (Galperin et al., 2019, table 13.1). In our experiment, and as discussed previously, we can estimate that the forcing scale is in fact about one to twice $L_\beta^\epsilon$. The forcing is thus already directly influenced by the $\beta$-effect, which is quite clear on the fluid response at the earliest times of our experiments (see Fig.9 in Lemasquerier et al. (2021)). The differences in the forcing mechanism and forcing scale may explain the rather non-universal estimates of $C_Z$ obtained in the past from simulations on the sphere, Cassini measurements of Jupiter's zonal flows and experiments. Second, it is probable that with a rather large forcing scale, we prevent the initial *isotropic* inverse energy cascade – an anisotropic cascade is present since the jets scale remains larger than the forcing scale. Since we do obtain the scaling expected in the zonostrophic regime, our experiments demonstrate that the scale of the forcing is not such an important parameter for the zonal flow to develop at large scales, and that a steep zonal spectra may develop even if $k_f \sim k_\beta$. To verify this hypothesis, we compute kinetic energy spectra from simulations QG2,3 and 4. We recall that for these simulations, we decrease the forcing scale by dividing the distance between vorticity sources by a factor 2 (see Table 4). The spectra corresponding to the steady state flows are represented in Fig.7(c-e). It is clear that as the forcing scale is reduced the scale separation between $k_f$ (beginning of the shaded area) and $k_\beta$ (intersection between the residual and zonal spectra) is increased, and a slope consistent with a -5/3 power law appears in that range.

To have a better scale separation between the forcing scale and the transitional scale, another possibility is to decrease the rotation rate, because it would decrease the free surface curvature, decrease the topographic $\beta$-effect, and hence increase $L_\beta$. We did not explore various rotation rates in the present study because the bottom plate was designed for a rotation rate of 75 RPM only. Second, one should keep in mind that when reducing the $\beta$-effect, the width and distance between the zonal jets will increase, and consequently finite-size effects might become too important. Third, decreasing the rotation rate will increase the Rossby and Ekman numbers, and the system will become less rotationally-constrained.

To sum up, the important conclusions of our spectral analysis are that:

1. Our experiments in Regime II are consistent with the picture of zonostrophic turbulence, both in terms of slope and amplitude of the theoretical spectra, which confirms their relevance to gas giants applications. This constitutes the first fully-experimental validation of the zonostrophic theory in a completely three-dimensional framework;

2. The absence of scale separation between $k_f$ and $k_\beta$ does not impede the establishment of a steep $k^{-5}$ slope for the zonal spectrum. Said differently, the zonal velocity profile obtained is not strongly sensitive to the forcing scale, at least in the range explored in the present study;

3. If one wishes to reach the regime of a well-developed isotropic inverse energy cascade, the goal should be to decrease the forcing scale, which is then a further experimental challenge.

## 5. Global and local potential vorticity mixing

We now turn to an analysis of the potential vorticity mixing, and use it to quantify turbulent dissipation rates independently of the spectral analysis presented in the previous section.

### 5.1. Staircasing

In the inviscid limit, the quasi-geostrophic model of the experimental flow reduces to the material conservation of the potential vorticity (PV) $q$ defined as

$$q(\rho, \phi, t) = \frac{\zeta(\rho, \phi, t) + f}{h(\rho)}. \tag{27}$$

As mentioned in the introduction, in the framework of PV mixing, narrow prograde jets correspond to strong gradients of PV whereas large retrograde jets coincide with regions of weak gradients, leading to the establishment of a PV staircase. In Fig.9(a,d), we plot instantaneous PV maps for the experiments J and A, which correspond to increasing Reynolds numbers. Because the fluid height $h$ increases exponentially with radius, the background PV profile, $f/h$, obtained when the flow is at rest in the rotating frame, is maximum at the center and decreases smoothly towards the edge of the tank. When zonal jets develop, the PV maps show that the PV does not decrease smoothly radially, but instead thin areas of sharp gradients develop. In Fig.9(a,d), we represent the associated time-averaged and azimuthally-averaged PV profiles, $\overline{q}_\phi(\rho)$. The steps are more readily visible on these profiles, and each of them corresponds to the presence of a prograde jet, whereas the regions in between are supporting a retrograde flow. Despite significantly increasing $Re$, and thus increasing the degree of mixing, the "staircasing" remains moderate even for our most extreme experiments.

Performing a Eulerian zonal average is the most straightforward way of obtaining the mean PV profile as a function of radius. This procedure works well when the jets are straight with no or weak meanders. However, Dritschel and Scott (2011) suggest that if the zonal jets are meandering, the zonal average can smear out the locally sharp PV gradients, and thereby underestimate the staircasing. Instead of performing a zonal Eulerian mean, the authors rather suggest to perform an average along the lines of constant PV to represent radial profiles of PV or velocity. Since our zonal jets are sometimes strongly meandering, especially at moderate forcing amplitude, we apply their procedure to compute PV profiles. The method is the following: PV contours are computed from an instantaneous PV map, then each contour is assigned an equivalent latitude (or radius). The equivalent radius for a given PV contour $\mathcal{C}$ is defined as the radius of the circular contour which encloses the same area as $\mathcal{C}$. We can then plot the PV of each contour as a function of the equivalent radius, denoted $q(\rho_{eq})$. This procedure was performed using codes developed by David Dritschel (personal communication). The corresponding results are plotted as pink lines on Fig.9(b,e). The insert on panel *(b)* shows that the PV gradients are only slightly stronger in the prograde jets with this method. Our conclusions are hence unchanged: the degree of staircasing is only moderate in our experiments, even in our most extreme cases.

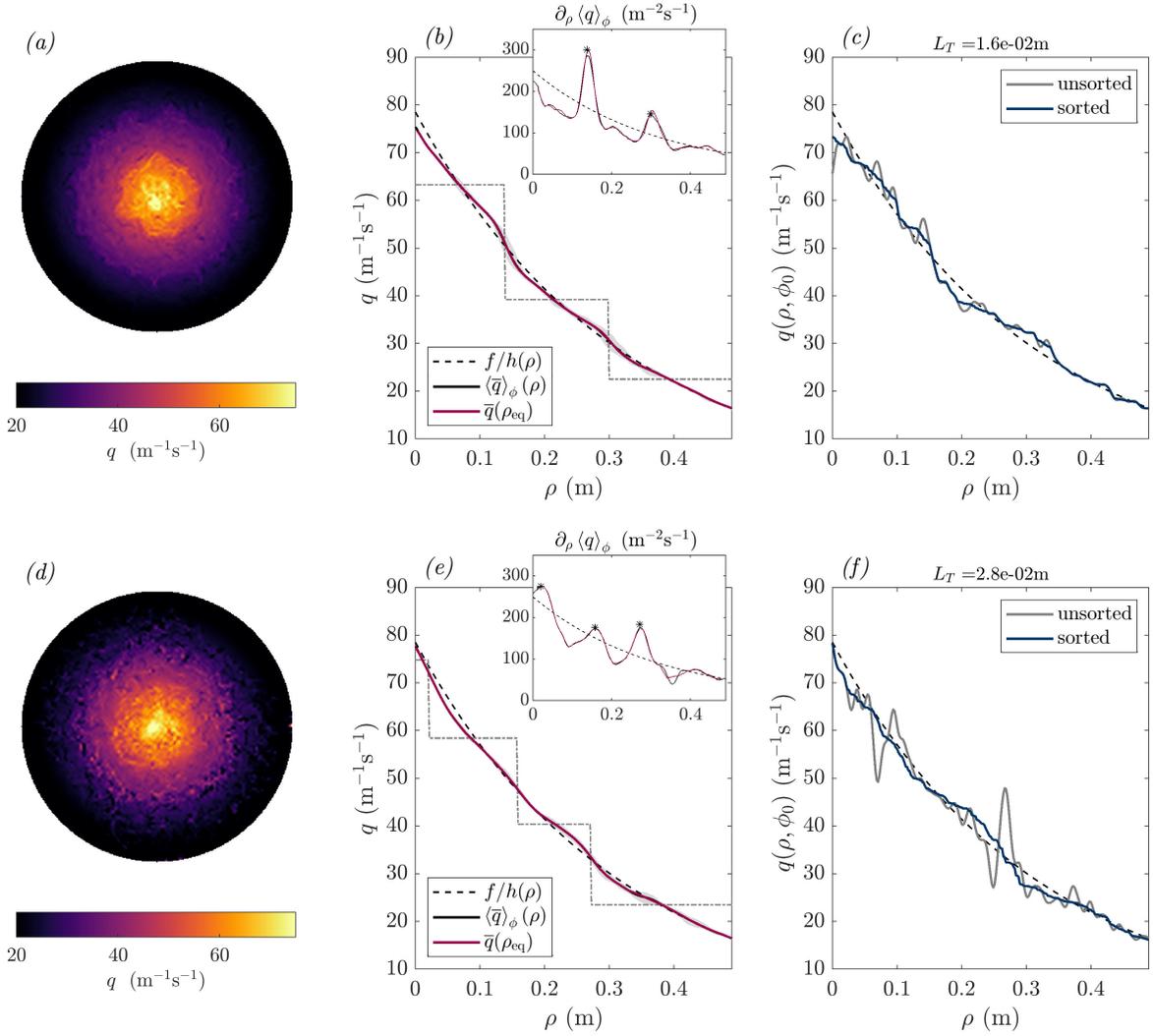

**Figure 9:** Illustration of potential vorticity mixing for experiments J (first row) and A (second row). *(a,d)* Instantaneous PV maps. *(b,e)* PV profiles. The dashed black line is the background PV profile, $f/h$, the continuous black line is the time and zonally-averaged PV profile $\langle \bar{q} \rangle_\phi(\rho)$, and the purple line is the equivalent latitude PV profile $\bar{q}(\rho_{eq}q)$. The grey lines show the variability of the PV profiles before performing the zonal average. The grey dashed-dotted line is the equivalent staircase $q_s$ defined in the text. The stars show the locations of PV jumps selected to compute the equivalent staircase. *(c,f)* Instantaneous PV profile at a fixed angular position $q(\rho, \phi_0)$. Grey line: before sorting, Blue line: after sorting. The Thorpe scale deduced from the sorting process for that specific angle $\phi_0$ is indicated at the top of each panel.

To provide a more quantitative estimate of the degree of staircaising, we use the method employed by Scott and Dritschel (2012). For each mean PV profile $\langle \overline{q} \rangle_\phi$, we define an equivalent staircase, $q_s$ such that the area below the curve is preserved, but with constant PV between each jump (see the dashed-dotted lines in Fig.9(b,e)). Note that the PV jumps are selected as local maxima in the PV radial derivative such that the difference between the real PV gradient and the background PV gradient is larger than an arbitrary value of 50 m$^{-2}$s$^{-1}$ (we use Matlab's *findpeaks* function). This threshold was chosen arbitrarily such that the identified peaks correspond to dynamically relevant jets, and not to small fluctuations in the zonal velocity profile (see the stars in Figures 10(d-f)). We then compute the integral quantities, $I_1$, $I_2$ and $I_3$ to quantify how far the actual profile is from the equivalent staircase and the background PV:

$$I_1 = \int_0^R |\langle \overline{q} \rangle_\phi - q_s| \, \mathrm{d}\rho,$$

$$I_2 = \int_0^R |q_s - f/h| \, \mathrm{d}\rho,$$

$$I_3 = \int_0^R |\langle \overline{q} \rangle_\phi - f/h| \, \mathrm{d}\rho.$$

The degree of staircaising can be evaluated by the quantities $I_{12} = 1 - I_1/I_2$ and $I_{23} = I_3/I_2$, which both take the value of 0 if $\langle \overline{q} \rangle_\phi = f/h$ (no staircaising) and 1 if $\langle \overline{q} \rangle_\phi = q_s$ (perfect staircaising). For the experiments J and A represented on Figure 9, we obtain $I_{12} = 0.14$ and $0.12$ and $I_{23} = 0.15$ and $0.16$ respectively. Based on Figure 13 in Scott and Dritschel (2012), these values would correspond to a zonostrophy index between approximately 3.5 and 4.5, which is consistent with the zonostrophy index estimated independently for our experiments. To achieve high degrees of staircaising, higher zonostrophy indices are needed, as well as vanishing forcing and vanishing dissipation. Even if strong, instantaneous jets form in our experiment, they are both strongly forced and strongly dissipated. The strong viscous friction constrains us to strongly force the flow in order to achieve sufficiently high Reynolds. It is hence not surprising that the degree of staircaising is weak in the experiments.

The QG simulations can again be employed to verify this hypothesis. Numerically, the viscous dissipation of the flow can be reduced by decreasing the Ekman number, everything else remaining unchanged (something that is not possible in the experiment). Simulations QG1b and QG1c have the exact same parameters as the reference simulation QG1a, but the Ekman number was reduced from $E_R = 1.2 \times 10^{-7}$ down to $6 \times 10^{-8}$ and $2 \times 10^{-8}$. Fig.10 shows instantaneous potential vorticity maps and time-averaged profiles for these three simulations. The inserts show that the slopes in the mean PV profile increase as the Ekman number is decreased. For the degree of staircaising, we obtain respectively $I_{12} = [0.17, 0.46, 0.38]$ and $I_{23} = [0.25, 0.54, 0.48]$. For the two simulations at smaller Ekman number, the staircases become more pronounced. Once again, these values are consistent with the results reported in Scott and Dritschel (2012). Note that the large dispersion of both $I_{12}$ and $I_{23}$ for the same zonostrophy index (see their Fig.13) can explain why the lowest Ekman case QG1c have a staircaising degree slightly smaller than the intermediate case QG1b.

## 5.2. Local mixing: the Thorpe scale

Even if strong, global staircaising does not develop, it is possible to quantify the *local* potential vorticity mixing in each experiment by measuring the equivalent of a mixing length, the so-called Thorpe scale. To this end, we follow the idea developed by Galperin et al. (2014a). We recall that the Thorpe scale is the rms displacement of fluid parcels when the PV profile is sorted into a monotonous profile. We apply this procedure to instantaneous PV profiles along the radius, for all the possible angles $\phi$. Examples of PV profiles before and after sorting are shown in Fig.9(c,f), and the resulting Thorpe scale is indicated at the top on each panel. The increase in the Thorpe scale and thus in the turbulent mixing is clear when the Reynolds number of the experiment is increased: experiment J (Fig.9(c)) shows only few deviations from a monotonous PV profile, meaning that overturns are not frequent and the PV is not efficiently mixed, whereas strong deviations are observed for experiment A (Fig.9(f)). The Thorpe scales measured for every experiment and numerical simulation are reported in Table 5.

The advantage of the Thorpe scale is that it is very easily measured a posteriori, for a given flow, and that it is an indirect measure of turbulence intensity (Thorpe, 2005). For stratified turbulence, with a buoyancy frequency $N$, the

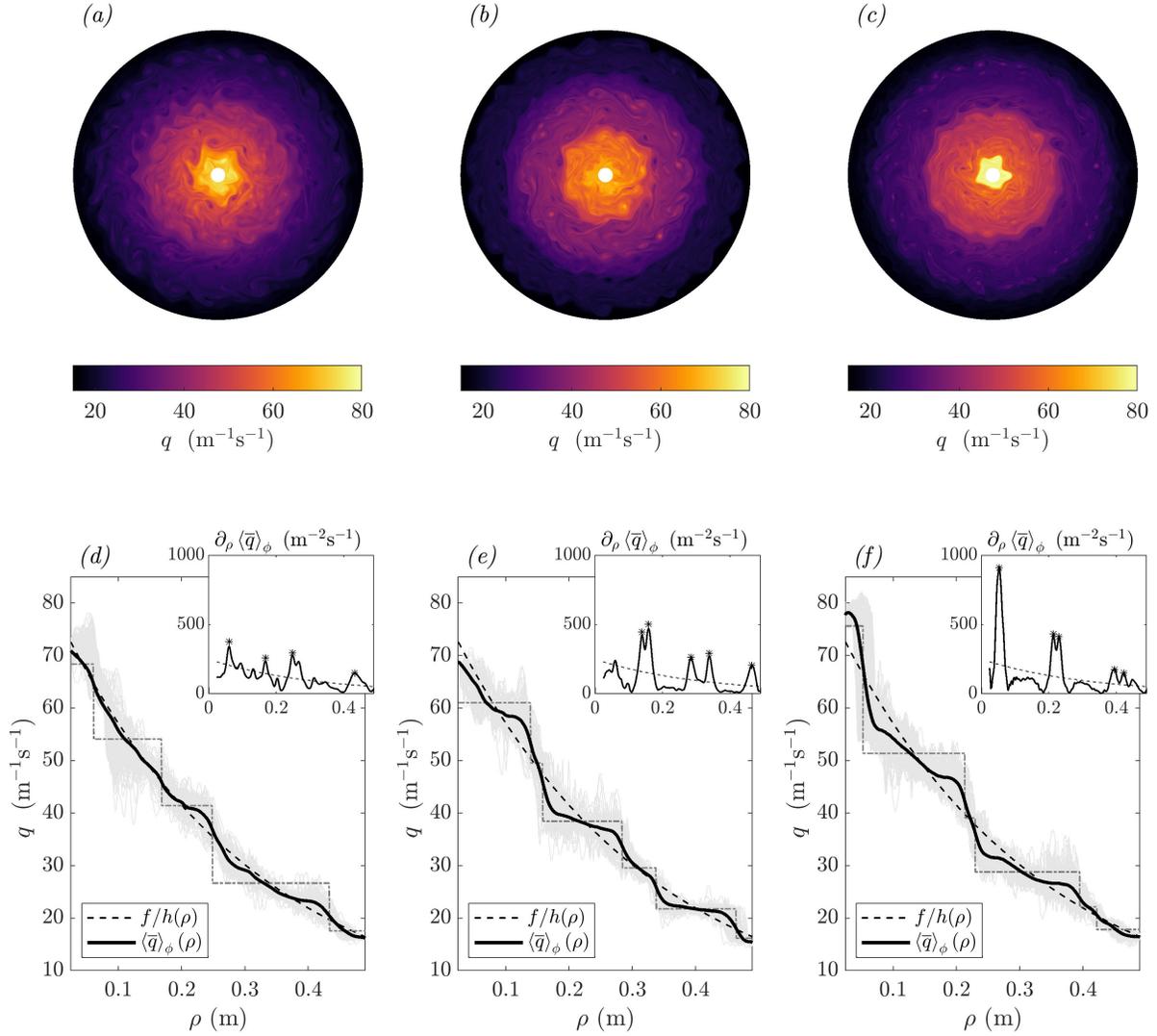

**Figure 10:** Potential vorticity maps and profiles in simulations QG1a *(a,d)*, QG1b *(b,e)* and QG1c *(c,f)*. *(a-c)* Instantaneous PV maps once in quasi-steady state. *(d-f)* PV profiles. The dashed black line is the background PV profile, $f/h$, the thick black line is the time and zonally-averaged potential vorticity profile $\langle \overline{q} \rangle_\phi(\rho)$. The grey lines show PV profiles for different angles before performing the zonal and time averages. The grey dashed-dotted line is the equivalent staircase $q_s$ defined in the text. The inserts show the radial derivative of the mean PV profile. The stars show the locations of PV jumps selected to compute the equivalent staircase.

Thorpe scale can be used to estimate the so-called Ozmidov scale, $L_O$ (Ozmidov, 1965; Thorpe and Deacon, 1977) which corresponds to the scale at which the turbulent turnover time is equal to the internal waves period, i.e. the scale at which the wave effect begins to dominate turbulence:

$$L_O = \left( \frac{\epsilon}{N^3} \right)^{1/2}. \tag{28}$$

For stratified turbulence, $L_O$ is thus the largest isotropic scale of the system, and is analogous to the transitional scale, $L_\beta$, for $\beta$-plane turbulence. Hence, if $L_T$ allows to estimate $L_O$ in stratified turbulence, it may allow to estimate $L_\beta$ in $\beta$-plane turbulence. This is of interest because $\epsilon$ can be estimated from $L_\beta$ (equation (3)).

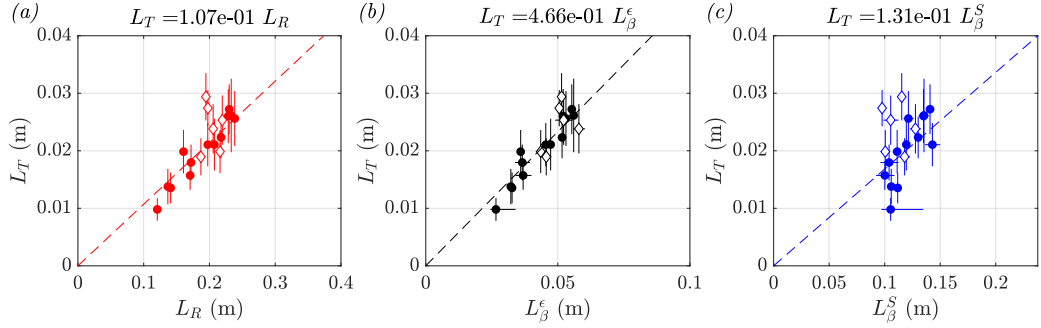

**Figure 11:** Comparison between the Thorpe scale $L_T$ and *(a)* the Rhines scale (equation (2)), *(b)* the transitional scale based on the upscale energy transfer, and *(c)* the spectral transitional scale (equation (26)). Circles: experiments. Diamonds: QG simulations. The equation at the top of each panel is the best linear fit of the experimental data, represented by the dashed line.

Here, we wish to confront this hypothesis to our experimental measurements, where we have measured $\epsilon$ and thus $L_\beta^\epsilon$ in a completely independent way using the spectral analysis. Fig.11 compares the Thorpe scale measured in our experiments and QG simulations with the Rhines scale $L_R$ and the transitional scales $L_\beta^\epsilon$ and $L_\beta^S$. These figures show that there is a very good correlation between $L_T$ and both $L_R$ and $L_\beta$, which holds for the whole range of $Re$ explored:

$$L_T \approx a L_\beta^\epsilon, \quad L_T \approx b L_R, \quad L_T \approx c L_\beta^S, \tag{29}$$

The fact that the correlation works equally well for those three scales arises from the fact that $L_R \sim R_\beta^\epsilon L_\beta^\epsilon$, and the zonostrophy index does not vary sufficiently for the two scales to behave independently. In our experimental set, $R_\beta^\epsilon \approx 3.69 - 4.71$, hence, we should have $a \approx 3.69 - 4.71\ b$, which is indeed what is observed. Similarly, $L_\beta^\epsilon = (C_Z/C_K)^{3/10} L_\beta^S$, with $(C_Z/C_K)^{3/10} \approx 0.29 - 0.41$. Hence, we expect $c \approx (C_Z/C_K)^{3/10} a$, which is again what is observed. That being said, considering $L_\beta^\epsilon$ alone, this result is a demonstration that $L_T$ can be used as a proxy of the typical length scale of the flow. If one assumes $L_T \sim 0.47 L_\beta^\epsilon$ as determined from Fig.11*(b)*, we can define a Thorpe estimate of the turbulent dissipation rate $\epsilon^T$ computed from the relation between $L_\beta$ and $\epsilon$ (equation (3)). The comparison between the measured $\epsilon^S$ and $\epsilon^T$ is represented in Fig.8*(b)*. There is a non-negligible dispersion around the 1:1 relationship, but the overall tendency is good, suggesting that PV sorting can effectively be used to estimate turbulent energy transfers.

## 6. Conclusions and Discussion

We have presented experimental results from a laboratory setup designed to reach extreme regimes of zonal jets relevant to gas giants (Cabanes et al., 2017; Lemasquerier et al., 2021). Two regimes of turbulent zonal jets have been identified by Lemasquerier et al. (2021). Here, we analyzed the statistical properties of the intense, experimental $\beta$-plane turbulence obtained in the super-resonant Regime II, for large forcing amplitudes. We showed that strong, instantaneous, turbulent zonal jets develop and contain up to 70% of the total kinetic energy of the flow in the most extreme experiments. These experiments are complemented by 2D quasi-geostrophic numerical simulations which allow us to reach the extreme experimental conditions in terms of Reynolds and Ekman numbers, and complement them by addressing various forcing and even more extreme regimes.

### 6.1. Zonostrophic turbulence and potential vorticity mixing

From a spectral analysis of the flow, we showed that the measured turbulent flow shares the properties of the so-called zonostrophic turbulence, relevant to the gas giants, in which the zonal flow alone contains more kinetic energy than the remaining of the flow. Along with Cabanes et al. (2017), these are the first experiments able to reach such extreme regimes (see Fig.1). While we retrieve the predicted steep slope for the zonal spectrum, it is hard to argue that a clear inverse cascade of energy develop in the experiments. Thanks to QG simulations, we show that this is due to the relatively large forcing scale of our experiment, which is in particular too close to the transitional scale. In other words, the forced eddies are directly influenced by the $\beta$-effect, and no isotropic inverse cascade of energy can robustly develop. Yet, this absence of inverse cascade does not impede the flow from feeding larger scales in an anisotropic way (i.e. feeding zonal structures), as expected for scales larger than the transitional scale. Interestingly, the forcing scale does not seem to have a significant influence on the final zonal jets profile obtained once in the quasi-steady state. This is consistent with the conclusions of Scott and Dritschel (2019), obtained in idealized simulations of potential vorticity mixing, and means that the relatively large forcing scale is not a significant limitation of our current experimental setup to study turbulent zonal jets.

At this point, we would like to raise the question of the "locality" of the zonal jets forcing. In a first paper (Lemasquerier et al., 2021), we showed that the jets developing in our experiment are initially due to wave-mean flow interactions, and emerge via the transport and deposition of momentum by Rossby waves, i.e. a streaming process. If this is particularly relevant for Regime I, not described here, the nature of the transition towards the second regime, a Rossby wave resonance, leads us to the hypothesis that wave-mean flow interactions are probably also of importance in the second regime, even in the highly turbulent cases. The acceleration of the zonal flow by a streaming process is non-local, and implies a direct transfer between waves and the mean flow. This picture, where both local cascades and nonlocal direct transfers coexist, is in line with quantifications of energy transfer in zonostrophic turbulence simulations, where both processes are at play (Galperin et al., 2019).

In terms of potential vorticity mixing, despite the strength and rectilinear shape of the zonal jets in our most extreme experiments, we showed that they are not accompanied by a strong potential vorticity mixing. This is because the zonostrophy index of our experiments is still "moderate" relatively to the cases leading to strong staircaising (Scott and Dritschel, 2012). Once again, QG simulations show that this is due to experimental limitations, since decreasing the viscous friction allow to reach higher degrees of staircaising. This result is important by itself because it bridges the gap between experiments and asymptotic numerical models. Furthermore, it shows that it is not necessary to be in the asymptotic regime of vanishing friction and forcing to obtain dominant turbulent zonal jets. In other words, strong, instantaneous zonal jets can be obtained even if the background PV gradient is moderately mixed, i.e. even if the relative vorticity associated with the jets is small compared to the background, planetary vorticity. Note that PV mixing studies (e.g. Dritschel and McIntyre, 2008) showed that there is a relation between the jets spacing (the staircase width) and the jets intensity (the PV jump at the interfaces). We could not test this relation in our experiments, because we are confronted with spatial confinement, and because only a small number of jets develops: the measure of the jets spacing is then compromised by finite-size effects. Next, we showed that it is possible to estimate the local intensity of the potential vorticity mixing by measuring the equivalent of the Thorpe scale (Thorpe, 2005). We verified experimentally that the measured Thorpe scale is proportional to the transitional scale, and that this relation can be used to deduce the upscale energy transfer rate. We showed that we obtain a good estimate of the upscale energy flux measured using spectral analysis. This result underlines the consistency between the zonostrophic turbulence and potential vorticity mixing theories.

## 6.2. Advantages and drawbacks of different methods for estimating the upscale turbulent energy transfer rate

Our experiments allow to address how efficiently kinetic energy spectra and PV mixing can be used to quantify the upscale energy transfer rate $\epsilon$ of a given turbulent flow with a $\beta$-effect. In the present paper, we have used three different methods:

1. The first and simplest estimate that can be made is based on the assumption that, in the experiment, the energy is dissipated by Ekman friction at a rate $\alpha = \Omega E^{1/2}$, and hence $\epsilon$ can be estimated simply by measuring the total kinetic energy of the flow: $\epsilon^E \sim u_{\mathrm{rms}}^2 \alpha/2$.

2. The second method, and perhaps the most rigorous one, consists in computing the residual kinetic energy spectra and fitting the $-5/3$ slope (equation (20)), $s$, which is proportional to $\epsilon^{2/3}$: $\epsilon^S \sim (s/C_K)^{3/2}$.

3. The third method consists in measuring a Thorpe scale by sorting local and instantaneous PV profiles. The deduced scale, assuming that it is correlated with $L_\beta$, allows to retrieve $\epsilon$ thanks to the relation $\epsilon^T \sim \beta^3 (L_\beta^T)^5$.

Method 1 gives a good order of magnitude for $\epsilon$, but we did not find an exactly linear relationship between $\epsilon^E$ and $\epsilon^S$ (Fig.8(a)), probably because bulk viscous dissipation and side friction are not completely negligible in our experiment. In addition, the main dissipation mechanism in natural flows is far from being a simple Ekman friction, and this is rather a practical scaling for idealized experimental or numerical model than a true generic method. While this first method assumes that dissipation is due to bottom viscous boundary layers only, the two other methods do not make any hypothesis on the nature of the dissipation process. Method 2, which we consider as the reference one, is probably the most robust but requires to have high resolution spatial and temporal observations to be able to compute energy spectra. For method 3, we show that the Thorpe scale is strongly correlated with the transitional scale, suggesting that the turbulent transfer rate could be efficiently retrieved in natural flows using potentially simpler measurements of instantaneous potential vorticity profiles. The success of this method however requires a precise estimate of the scaling prefactor for the relationship between $L_T$ and $L_\beta$. Here, we found $L_\beta \sim 2.1 L_T$, but this factor may potentially vary. We show nevertheless that this scaling holds for the whole range of Reynolds numbers explored. A complete systematic study, where $E$ and $R_\beta$ are varied independently would allow to determine the validity of this scaling depending on the flow regime. Note that the Thorpe scale is routinely used in oceanography to measure the vertical mixing of stratified turbulence. Our results support the analogy raised by Galperin et al. (2014a), i.e. that an equivalent method can be used to measure the lateral mixing due to $\beta$-plane turbulence, as done by Cabanes et al. (2020) for Jupiter and Saturn.

### 6.3. Implication for Jupiter's cloud layer dynamics

What are the implications of our results regarding the dynamics observed in the cloud layer of Jupiter? We recall that the theory of zonostrophic turbulence was developed in a purely two-dimensional framework, thanks to numerical simulations on the sphere, i.e. indirectly assuming that Jupiter's turbulence is shallow and confined in its weather layer. It is thus not obvious a priori that this theory is still relevant in the case of Jupiter and Saturn jets, which are now strongly believed to be deep, extending as deep as 3,000 and 9,000 km respectively (Kaspi et al., 2018; Galanti et al., 2019; Kaspi, Galanti, Showman, Stevenson, Guillot, Iess and Bolton, 2020). Quantitative analyses of the Cassini 70-days movie of Jupiter's clouds support the picture of an inverse energy cascade in Jupiter's weather layer (Choi and Showman, 2011; Galperin et al., 2014b). Using the same data, Young and Read (2017) showed that there is indeed an inverse transfer of energy from scales of 2,000-3,000 km, up to 20,000 km. However, they also identify a direct transfer of energy from 2,000 km towards small scales, which is not expected in classical 2D turbulence. In addition, they show that most energy transfer occurs between eddies and the zonal flow directly, i.e. non-locally. The scale of transition between direct and inverse energy transfer is close to the first Rossby radius of deformation, which supports the idea that the forcing is due to baroclinic instabilities. However, one should keep in mind that the velocity fields from which these conclusions are drawn are measured in the weather layer only. It is possible that the forcing of the jets at depth occurs via small scale convective motions, while baroclinic processes dominate in the shallow weather layer, with coupling mechanisms unexplored to date. Note that dual cascades to small scales and large scales have been reported for several physical systems, such as magnetohydrodynamics in two and three space dimensions, surface capillary waves and rotating-stratified turbulence (see Pouquet, Marino, Mininni and Rosenberg (2017) for an overview, and Galperin and Sukoriansky (2020)).

In the present experimental setup, the zonal jets are barotropic jets extending parallel to the rotation axis in a deep layer of water, which is hence representative of the deep jets configuration contrary to previous shallow water experiments. Our setup allows to model a fully three-dimensional system, without any *a priori* assumption on the two-dimensionality of the flow. The fact that jets can emerge spontaneously from a deep, fast rotating, fully-3D turbulent flow, even in the presence of a large bottom drag (relatively to that on Jupiter), supports the deep hypothesis of zonal jets origin. Furthermore, the spectral analysis shows that, even for deep jets, the spectral properties derived in the framework of two-dimensional turbulence can be recovered. Therefore, the fact that these scalings are also measured on Jupiter (Choi and Showman, 2011; Galperin et al., 2014b) does not allow to discriminate between shallow and deep jets. One possible line of investigation would be to continue the spectral analysis to confirm or not the presence of a turbulent inverse cascade of energy by computing structure functions. This would also allow for a direct comparison with the aforementioned turbulent statistics of Jupiter, and in particular, we could verify if we also retrieve a direct energy cascade despite the quasi-two-dimensionality of our fast-rotating experimental flows. This analysis was out of reach with our current experimental datasets because better time and space-resolved PIV measurements are required. Finally we wish to conclude by underlining that if Jupiter's zonal jets are deep and emerge from deep thermal convection, it is possible that the turbulence observed in the cloud layer is somehow "superimposed" to a deeper turbulence from which the zonal jets emerge. In such a case, comparing the residual and zonal kinetic energy spectra should be taken with caution given that we may be looking at the superposition or coexistence of flows having different physical origins.

As already mentioned in the introduction, we want to underline that exploring barotropic experimental or numerical models is still relevant for systems which exhibit a predominantly baroclinic circulation, such as oceans and atmospheres. Indeed, planetary, quasi-geostrophic flows can undergo a phenomenon of *barotropization*. This process refers to the fact that following baroclinic instabilities, the energy initially cascades from the scale of meridional temperature gradients down to the radius of deformation (commensurate with the most unstable wavenumber). There is subsequently a conversion to barotropic energy, because the energy tends to feed the gravest mode, which is a barotropic (vertically-invariant) mode. Finally, an inverse cascade to larger modes occur (Rhines, 1977; Salmon, 1978). Said differently, in QG flows, the energy cascade to larger horizontal scales is generally accompanied by a cascade to larger vertical scales, and the barotropic modes can end up containing a significant fraction of the energy (see Charney (1971) and Chapter 12 in Vallis (2006)). For the Earth, both GCM simulations and reanalysis of atmospheric and oceanic data confirmed this tendency to strong barotropization (see Galperin et al., 2019, and references therein). In baroclinic flows, the barotropization occurs at a scale close to the internal radius of deformation, which should be small enough to leave room for an inverse cascade in the barotropic modes. In our experiment, instead of relying on baroclinic instabilities to inject energy, we directly (mechanically) force the flow at a small and well-controlled scale, and the barotropic flow grows by itself because of the quasi-geostrophy of the system.

Finally, in the present manuscript, we focus on gas giants dynamics at midlatitudes, where the large variation of the Coriolis force with radius is responsible for a strong $\beta$-effect and hence strong zonal flows. At increasing latitudes, the gradient of the Coriolis force decreases and tends towards zero at the pole. Consequently, the zonal jets weaken, and the dynamics becomes more isotropic, with a predominance of vortices rather than jets. In addition, the vortices become predominantly cyclonic, contrary to what is observed at midlatitudes. This transition in the dynamics with increasing latitude has been first suggested by Theiss (2004) for Earth's oceans, and later observed in shallow-water turbulence models on the sphere (e.g. Cho and Polvani (1996); Scott and Polvani (2007)). When the Juno spacecraft arrived at Jupiter, visible and infrared observations from above the poles revealed an incredible dynamics consisting in persistent polygonal patterns of cyclones around both poles (Adriani, Mura, Orton, Hansen, Altieri, Moriconi, Rogers, Eichstädt, Momary, Ingersoll, Filacchione, Sindoni, Tabataba-Vakili, Dinelli, Fabiano, Bolton, Connerney, Atreya, Lunine, Tosi, Migliorini, Grassi, Piccioni, Noschese, Cicchetti, Plainaki, Olivieri, O'Neill, Turrini, Stefani, Sordini and Amoroso, 2018). A natural extension of this work is hence to look at how the dynamics evolve for a progressively less dominant $\beta$-effect.

## 6.4. Advantages and limitations of experimental approaches of zonal jets

We wish to conclude this study by a broader discussion on the approach and methods employed in the prospect of better understanding gas giants dynamics. The regimes reached experimentally ($E \sim 3 \times 10^{-7}$, $Re \sim 10^4$, $Ro \sim 10^{-3}$) are not impossible to reach in direct numerical simulations, but at very large cost. To give an order of magnitude, a single DNS of 1,000 rotation times at an Ekman number three times larger than in the experiment ($E \sim 10^{-6}$) represents 13 minutes of an experiment, but would require 13 days of computation on 2048 CPU cores (650,000 CPU hours). This type of DNS was performed in Cabanes et al. (2017), but it is clear that numerical systematic studies in these regimes are inconceivable. The high cost of these DNS is inherent to 3D simulations of geostrophic turbulent flows, where both large-scale structures and small-scale turbulent eddies and inertial waves are present and need to be resolved simultaneously. On the contrary, one experimental realisation "costs" about 2 days (one for the actual experiment, and one for saving and post-processing the images through the PIV algorithm), which allows for multiple realisations and exploration of the parameter space. In addition, the dynamics of the large-scale jets is slow, and results from cumulative effects from the underlying turbulence. Studying their long-term dynamics requires to wait for very long times. Experimentally, we can easily reach several thousands of rotation times, and time-resolved particle image velocimetry allows for high-resolution records of the interactions between the turbulent flow and the slowly evolving large-scale jets.

Of course, idealized numerical models allow to circumvent these difficulties. This is the case for instance of reduced two-dimensional models, such as shallow-water models and quasi-geostrophic models including the one used in the present study. Statistical simulations and quasilinear models (e.g. Constantinou, Farrell and Ioannou, 2014) where eddy-eddy interactions are neglected can also be used. Recently, rare events algorithm have also been employed to study multistability and spontaneous transitions among zonal jets (Bouchet, Rolland and Simonnet, 2019). However, the assumptions underlying each of these models and their relevance for gas giants dynamics need to be systematically addressed.

Despite their clear advantages, experiments also come with their own limitations, as underlined throughout this paper:

- The spatial confinement of the experiment is problematic for studying zonal jets equilibration. For instance, finite-size effects "discretize" the evolution of jets spacing when varying any control parameter. Due to confinement, jets are not free to evolve in space, and this may for instance impede long-term drift or nucleation of jets. One could nevertheless argue that jets on the planets are also confined, but there are still about ten prograde jets on Jupiter, leading to a scale separation between the size of the domain and the jets. We should hence seek to build experiments where the scale of the tank is large compared to the scale of the jets, which itself is large compared to the scale of injection. This is challenging, because when trying to reach more extreme regimes, we increase the turbulence intensity or decrease the Ekman friction, which increases the jets scale. Imposing a fast rotation on larger scale containers is also a technical challenge.

- Experiments at high rotation rates allow to reach regimes with small friction (here $E \sim 3 \times 10^{-7}$), but never asymptotically small. We stand in a strongly forced-dissipative regime, and it is difficult to bridge the gap with idealized models where both the forcing and the dissipation are vanishing, as discussed in the case of potential vorticity mixing.

- The experimental forcing is performed at small-scale, but it is difficult to reach the same scale separation as in numerical simulations and less straightforward to change the forcing properties. Exploring numerically the effect of the forcing properties (scale, spatio-temporal stationarity...) is hence required to extrapolate experimental results to more realistic planetary conditions, as proposed here using QG simulations.

- Finally, a last difficulty is that we cannot vary independently the Ekman and Reynolds numbers and the zonostrophy index of the flow. Changing the rotation rate modifies the Ekman number but also the zonostrophy index because it changes the free surface shape. One way of avoiding this is to use a sloping bottom and a rigid lid instead of a free surface for the $\beta$-effect, but it introduces a supplementary friction which can kill the zonal flow.

These limitations justify that we employed idealized numerical simulations to reproduce the experimental conditions and then explore one by one some effects artificially introduced by the experimental constraints (forcing nature, spatial confinement, high viscosity, boundary conditions, etc). Note that it is also important to consider the addition of physical effects which will hardly be incorporated in experiments, such as magneto-hydrodynamical dissipation of zonal flows or compressibility effects.

Finally, let us underline that the goal of the idealized experimental and numerical models presented here is to shed light on fundamental mechanisms governing the dynamics of the system, rather than be quantitatively predictive regarding the specific case of Jupiter. Jupiter exhibits a wealth of dynamical processes, occurring at very different temporal and spatial scales. Our understanding of Jupiter and the gas giants as global systems is still complicated by numerous sources of uncertainty and technical limitations. The continuous improvement of technical and computational capabilities allows experimental and numerical models to get closer to the planetary regimes. However, understanding Jupiter by forward modelling requires observations of sufficiently good quality and coverage to which the models outputs can be compared. The recent, accurate observations of Jupiter and Saturn from the Juno and Cassini missions constitute a very good opportunity in this regard, while introducing new and exciting challenges for planetary modellers at the same time.

## Aknowledgements

The authors acknowledge funding by the European Research Council under the European Union's Horizon 2020 research and innovation program through Grant No. 681835-FLUDYCO-ERC-2015-CoG. The authors are most grateful to E. Bertrand and W. Le Coz for their help during the conception and building of the experiment, and to J.-J. Lasserre for his help in setting up the PIV system. We also thank David Dritschel for providing numerical codes to compute equivalent latitude potential vorticity profiles. *Centre de Calcul Intensif d'Aix-Marseille* is acknowledged for granting access to its high-performance computing resources. This work was performed using HPC resources from GENCI–IDRIS (Grants 2021-A0100407543 and 2022-A0120407543).

## Data Availability

The data used to plot Figures 1, 8 and 11 are provided in Tables 3, 5 and B.

## Declaration of interests

The authors declare that they have no known competing financial interests or personal relationships that could have appeared to influence the work reported in this paper.

## A. Hovmoller diagrams and zonal flows for experiment and QG simulations

On Figure 12, we plot space-time diagrams and zonal flow profiles for a typical experiment and four QG numerical simulations.

## B. Data from previous experiments

In Table B, we provide the dimensional and non-dimensional parameters that are used to plot Fig.1, where the regimes reached by experiments on zonal jets are compared. Note that various forcing types are employed. The definition of the Ekman number $E$, Reynolds number $Re$ and and zonostrophy index $R_\beta$ are those of Table 1.

## C. Quasi-geostrophic model of the experiment

### C.1. Derivation

We use the cylindrical coordinates $(\rho, \phi, z)$ with $z$ oriented downward and $(\mathbf{e}_\rho, \mathbf{e}_\phi, \mathbf{e}_z)$ the associated unit vectors (Fig. 2). We consider the flow of an incompressible fluid of constant kinematic viscosity $\nu$ and density $\rho_f$, rotating around the vertical axis at a constant rate $\mathbf{\Omega} = \Omega \, \mathbf{e}_z$. In our experimental setup, $\Omega > 0$ since the turntable rotates in the clockwise direction. We denote the velocity field $\mathbf{u} = (u_\rho, u_\phi, u_z)_{\mathbf{e}_\rho, \mathbf{e}_\phi, \mathbf{e}_z}$. The fluid is enclosed inside a cylinder of radius $R$. The lower boundary is a rigid plate located at $z = 0$ and the upper boundary is a free surface defined by $z = -h(\rho)$. Note that here we assume that our experiment, which has a parabolic free-surface and a curved bottom, can be modeled with a flat bottom and an exponential free-surface. Doing so, we neglect the influence of the shape of the bottom topography on the vertical velocity (see equation (37)). The experimental parameters were carefully chosen such that the bottom topography is as small as possible in amplitude (resulting in a maximum height difference of 5.36 cm and a mean absolute slope of 22%). Thus, one should keep in mind that the presently derived model is only valid for relatively smooth bottom topographies for which we can use the expression of the Ekman pumping over a flat surface. If the topography was of high amplitude, then the local inclination of the bottom boundary would enter the QG model because it modifies the Ekman pumping (Greenspan, 1968).

We start from the continuity and horizontal Navier-Stokes equations in the rotating frame:

$$\frac{\partial u_\rho}{\partial t} + u_\rho \frac{\partial u_\rho}{\partial \rho} + \frac{u_\phi}{\rho} \frac{\partial u_\rho}{\partial \phi} - \frac{u_\phi^2}{\rho} - f u_\phi = -\frac{1}{\rho_f} \frac{\partial P}{\partial \rho} + \nu \left( \nabla^2 u_\rho - \frac{u_\rho}{\rho^2} - \frac{2}{\rho^2} \frac{\partial u_\phi}{\partial \phi} \right), \tag{30}$$

$$\frac{\partial u_\phi}{\partial t} + u_\rho \frac{\partial u_\phi}{\partial \rho} + \frac{u_\phi}{\rho} \frac{\partial u_\phi}{\partial \phi} + \frac{u_\phi u_\rho}{\rho} + f u_\rho = -\frac{1}{\rho_f} \frac{1}{\rho} \frac{\partial P}{\partial \phi} + \nu \left( \nabla^2 u_\phi - \frac{u_\phi}{\rho^2} + \frac{2}{\rho^2} \frac{\partial u_\rho}{\partial \phi} \right), \tag{31}$$

$$\frac{1}{\rho} \frac{\partial (\rho u_\rho)}{\partial \rho} + \frac{1}{\rho} \frac{\partial u_\phi}{\partial \phi} + \frac{\partial u_z}{\partial z} = 0, \tag{32}$$

where $\nabla^2 \cdot = \partial_\rho^2 \cdot + \partial_\phi^2 \cdot /\rho^2 + \partial_\rho \cdot /\rho$. The Coriolis parameter is $f = 2\Omega$ and $P = p + \rho_f g z - \rho_f f^2 \rho^2/8$ is the reduced pressure incorporating the gravity and centrifugal effects. Note that if we neglect the vertical dependence of the horizontal velocity, we keep it for the vertical velocity $u_z$. Indeed, as previously explained, $u_z$ is expected to strongly vary close to the top and bottom boundaries, and we want to take into account these effects on the horizontal velocity divergence.

The curl of the Navier-Stokes equation leads to the vorticity equation

$$\frac{\partial \zeta}{\partial t} + u_\rho \frac{\partial \zeta}{\partial \rho} + \frac{u_\phi}{\rho} \frac{\partial \zeta}{\partial \phi} + (\zeta + f) \, \nabla_h \cdot \mathbf{u} = \nu \nabla^2 \zeta, \tag{33}$$

where $\zeta = (\nabla \times \mathbf{u}) \cdot \mathbf{e}_z = (\partial_\rho(\rho u_\phi) - \partial_\phi u_\rho)/\rho$ is the vertical component of the vorticity and $\nabla_h \cdot \mathbf{u}$ is the horizontal divergence

$$\nabla_h \cdot \mathbf{u} = \frac{1}{\rho} \frac{\partial (\rho u_\rho)}{\partial \rho} + \frac{1}{\rho} \frac{\partial u_\phi}{\partial \phi}. \tag{34}$$

The last term of the left hand side of equation (33), the vortex stretching term, involves the horizontal divergence of the flow which can be estimated from equation (32) after integration from $z = -h(\rho)$ to $z = 0$ ($z$ oriented downward) to unveil the Ekman pumping through the vertical velocity:

$$\boldsymbol{\nabla}_h \cdot \boldsymbol{u} = -\frac{1}{h(\rho)} \int_{z=-h}^{0} \frac{\partial u_z}{\partial z}\, \mathrm{d}z = \frac{u_z\big|_{z=-h} - u_z\big|_{z=0}}{h(\rho)}. \tag{35}$$

The vertical velocity at the free surface $u_z\big|_{z=-h}$ is given by the kinematic condition

$$u_z\big|_{z=-h} = -\left(\frac{\partial h}{\partial t} + u_\rho \frac{\partial h}{\partial \rho} + \frac{u_\phi}{\rho}\frac{\partial h}{\partial \phi}\right) = -u_\rho \frac{\partial h}{\partial \rho}, \tag{36}$$

since $h$ is axisymmetric and we neglect any temporal variations of the fluid height (rigid lid approximation). The vertical velocity at the bottom $u_z\big|_{z=0}$ results from the no-slip boundary condition generating an Ekman pumping. According to linear Ekman theory, for a flat bottom and small Rossby number, the vertical velocity at the top of the boundary layer is proportional to the relative vorticity in the interior flow (see section 5.7 in Vallis, 2017):

$$u_z\big|_{z=0} = -\frac{1}{2}\delta\zeta = -\frac{1}{2}E^{1/2}h_0\zeta, \tag{37}$$

where $\delta = \sqrt{2\nu/f}$ is the thickness of the Ekman layer and $E = 2\nu/(f h_0^2)$ is the Ekman number, $h_0$ being the mean fluid height. The horizontal divergence (35) is then

$$\nabla_h \cdot \mathbf{u} = -\frac{u_\rho}{h}\frac{\mathrm{d}h}{\mathrm{d}\rho} + \frac{E^{1/2}}{2}\frac{h_0}{h}\zeta. \tag{38}$$

The squeezing and stretching of vorticity is hence due to both the changes in the fluid depth and the vertical velocity induced by the Ekman boundary layer.

Substitution of the horizontal divergence (38) in the vorticity equation (33) yields

$$\frac{\partial \zeta}{\partial t} + u_\rho \frac{\partial \zeta}{\partial \rho} + \frac{u_\phi}{\rho}\frac{\partial \zeta}{\partial \phi} \underbrace{-(\zeta + f)\frac{u_\rho}{h}\frac{\mathrm{d}h}{\mathrm{d}\rho}}_{\text{Topographic }\beta-\text{effect}} + \underbrace{\frac{E^{1/2}}{2}\frac{h_0}{h}(\zeta + f)\zeta}_{\text{Ekman pumping}} = \nu\nabla^2\zeta, \tag{39}$$

We switch to non-dimensional variables using $1/f$ as the timescale and the radius of the tank, $R$, as the length-scale, and we denote the non-dimensional variables with a tilde such that

$$\begin{aligned}
\zeta &= \tilde{\zeta}f, \tag{40}\\
u_\rho &= \tilde{u}_\rho f R, \tag{41}\\
u_\phi &= \tilde{u}_\phi f R, \tag{42}\\
\rho &= \tilde{\rho}R, \tag{43}\\
t &= \tilde{t}/f, \tag{44}\\
h &= \tilde{h}R, \tag{45}
\end{aligned}$$

We keep the tilde in the following to better identify non-dimensional variables, and to avoid confusion with the experimental variables and parameters which are always given in dimensional forms first. Equation (39) becomes

$$\frac{\partial \tilde{\zeta}}{\partial \tilde{t}} + \tilde{u}_\rho \frac{\partial \tilde{\zeta}}{\partial \tilde{\rho}} + \frac{\tilde{u}_\phi}{\tilde{\rho}}\frac{\partial \tilde{\zeta}}{\partial \phi} - \frac{\tilde{u}_\rho}{\tilde{h}}\frac{\mathrm{d}\tilde{h}}{\mathrm{d}\tilde{\rho}}(\tilde{\zeta} + 1) + \frac{E_R^{1/2}}{2\tilde{h}}(\tilde{\zeta}+1)\tilde{\zeta} = \frac{E_R}{2}\tilde{\nabla}^2\tilde{\zeta} + \tilde{F}, \tag{46}$$

where $E_R$ is the Ekman number based on the radius of the tank, $E_R = 2\nu/(f R^2) = (h_0/R)^2 E$, and we have introduced a forcing term $\tilde{F}$. To close this equation, we now need an expression for the horizontal components of the velocity. To do so, we use the definition of $\zeta$ to rewrite the expression of the horizontal divergence (equation (38)) as a zero-divergence for a modified velocity field:

$$\frac{1}{\tilde{\rho}}\frac{\partial(\tilde{\rho}\tilde{u}_\rho)}{\partial\tilde{\rho}} + \frac{1}{\tilde{\rho}}\frac{\partial\tilde{u}_\phi}{\partial\phi} = -\frac{\tilde{u}_\rho}{\tilde{h}}\frac{\partial\tilde{h}}{\partial\tilde{\rho}} + \frac{E_R^{1/2}}{2\tilde{h}}\tilde{\zeta} \tag{47}$$

$$\implies \frac{\partial}{\partial\tilde{\rho}}\left(\tilde{\rho}\left[\tilde{h}\tilde{u}_\rho - \tilde{u}_\phi\frac{E_R^{1/2}}{2}\right]\right) + \frac{\partial}{\partial\phi}\left(\tilde{h}\tilde{u}_\phi + \tilde{u}_\rho\frac{E_R^{1/2}}{2}\right) = 0. \tag{48}$$

This allows us to define a streamfunction $\tilde{\psi}$ such that

$$\tilde{h}\tilde{u}_\rho - \tilde{u}_\phi\frac{E_R^{1/2}}{2} = \frac{1}{\tilde{\rho}}\frac{\partial\tilde{\psi}}{\partial\phi}, \tag{49}$$

$$\tilde{h}\tilde{u}_\phi + \tilde{u}_\rho\frac{E_R^{1/2}}{2} = -\frac{\partial\tilde{\psi}}{\partial\tilde{\rho}}, \tag{50}$$

or equivalently

$$\tilde{u}_\rho = \frac{1}{\tilde{h}}\frac{1}{1+E_R\tilde{h}^{-2}}\left(\frac{1}{\tilde{\rho}}\frac{\partial\tilde{\psi}}{\partial\phi} - \frac{E_R^{1/2}}{2\tilde{h}}\frac{\partial\tilde{\psi}}{\partial\tilde{\rho}}\right) = \frac{1}{\tilde{h}}\left(\frac{1}{\tilde{\rho}}\frac{\partial\tilde{\psi}}{\partial\phi} - \frac{E_R^{1/2}}{2\tilde{h}}\frac{\partial\tilde{\psi}}{\partial\tilde{\rho}}\right) + \mathcal{O}(E_R\tilde{h}^{-2}), \tag{51}$$

$$\tilde{u}_\phi = \frac{1}{\tilde{h}}\frac{1}{1+E_R\tilde{h}^{-2}}\left(-\frac{E_R^{1/2}}{2\tilde{h}\tilde{\rho}}\frac{\partial\tilde{\psi}}{\partial\phi} - \frac{\partial\tilde{\psi}}{\partial\tilde{\rho}}\right) = \frac{1}{\tilde{h}}\left(-\frac{E_R^{1/2}}{2\tilde{h}\tilde{\rho}}\frac{\partial\tilde{\psi}}{\partial\phi} - \frac{\partial\tilde{\psi}}{\partial\tilde{\rho}}\right) + \mathcal{O}(E_R\tilde{h}^{-2}), \tag{52}$$

where we have neglected terms of order greater or equal to $\mathcal{O}(E_R\tilde{h}^{-2})$, which is justified in the limit where we stand since $\tilde{h}$ is of order unity, and $E_R \ll 1$. Physically, this approximation means that the Ekman boundary layers are very thin compared to the fluid height. Substituting the horizontal velocities with their expressions (51) and (52) into the vorticity equation (46), we obtain in its condensed form the final vorticity equation

$$\frac{\partial\tilde{\zeta}}{\partial\tilde{t}} + \mathcal{J}(\tilde{q},\tilde{\psi}) - \frac{E_R^{1/2}}{2\tilde{h}}\tilde{\nabla}\tilde{\psi}\cdot\tilde{\nabla}\tilde{q} = \frac{E_R}{2}\tilde{\nabla}^2\zeta - \frac{E_R^{1/2}}{2\tilde{h}}\tilde{\zeta}(\tilde{\zeta}+1) + \tilde{F} \tag{53}$$

where $\mathcal{J}$ is the non-dimensional Jacobian operator in cylindrical coordinates

$$\mathcal{J}(a,b) = \frac{1}{\tilde{\rho}}\left(\frac{\partial a}{\partial\tilde{\rho}}\frac{\partial b}{\partial\phi} - \frac{\partial b}{\partial\tilde{\rho}}\frac{\partial a}{\partial\phi}\right), \tag{54}$$

and we introduced the potential vorticity

$$\tilde{q} = \frac{\tilde{\zeta}+1}{\tilde{h}}. \tag{55}$$

Taking the curl of (51)-(52), we obtain the modified Poisson equation that links the vorticity and the streamfunction, which closes the system of equations:

$$\tilde{\zeta} = -\frac{1}{\tilde{h}}\tilde{\nabla}^2\tilde{\psi} + \frac{1}{\tilde{h}^2}\tilde{\nabla}\tilde{h}\cdot\tilde{\nabla}\tilde{\psi} + \frac{E_R^{1/2}}{\tilde{h}^2}\mathcal{J}(\tilde{h},\tilde{\psi}). \tag{56}$$

Note that the potential vorticity $\tilde{q}$ is a materially conserved quantity if we drop the forcing term and neglect viscous effects. Equation (53) can indeed be recast as

$$\frac{\partial\tilde{q}}{\partial\tilde{t}} + \frac{1}{\tilde{h}\tilde{\rho}}\left(\frac{\partial\tilde{\psi}}{\partial\phi}\frac{\partial\tilde{q}}{\partial\tilde{\rho}} - \frac{\partial\tilde{\psi}}{\partial\tilde{\rho}}\frac{\partial\tilde{q}}{\partial\phi}\right) = 0 \tag{57}$$

$$\Leftrightarrow \frac{D\tilde{q}}{D\tilde{t}} = 0. \tag{58}$$

## C.2. Justification for keeping higher order, non-linear terms

In the case of QG models derived for rotating convection (e.g. Cardin and Olson, 1994; Aubert et al., 2003; Gillet and Jones, 2006), the Ekman pumping effects are incorporated the same way as we did, except that their geometry is more complicated and the no-slip boundaries cannot be considered as flat. Apart from these geometrical factors, the main difference lies in the fact that in (33), it is common in the rotating convection community to assume that the local vorticity $\zeta$ is negligible compared to the planetary vorticity $f$, and retain only the linear terms for the topographic $\beta$-effect and Ekman pumping. In that case, the terms $E_R^{1/2}/2\tilde{h} \, \nabla\tilde{\psi} \cdot \nabla\tilde{q}$ and $\tilde{\zeta}^2 E_R^{1/2}/2\tilde{h}$ are removed from equation (53). These two terms hence represent corrections due to nonlinear Ekman effects. The first term corrects the potential vorticity advection, while the second one is a correction of vortex stretching effects. Following Sansón and Van Heijst (2000, 2002), we argue that these terms should be kept in our simulations since they are at least of same order as the term $E_R/2\nabla^2\tilde{\zeta}$ which represents the bulk viscous effects.

This can be verified by introducing the local Rossby number, based on the local vorticity, $Ro_\zeta = \zeta/f = \tilde{\zeta}$, which is different from the global Rossby number defined in the main text, $Ro = u_{\mathrm{rms}}/fR$. In our case, we are in a regime where both the Ekman and global Rossby numbers are small, meaning that rotation dominates respectively viscous effects and inertia. Both conditions are mandatory to legitimately assume a two-dimensionalization of the flow. However, the local Rossby number may not be small and the local vorticity $\zeta$ associated with small turbulent eddies may be of same order as the rotation rate $f/2$, which is indeed verified in our simulations. This justifies the fact that we keep supplementary non-linear terms compared to other studies. Using the (non-dimensional) Ekman spin-down time scale $E^{-1/2}$ as the reference time scale, the different terms of equation (53) are indeed of order

$$a. \ \frac{\partial \tilde{\zeta}}{\partial \tilde{t}} \sim Ro_\zeta E^{1/2} \sim 3 \times 10^{-5}, \tag{59}$$

$$b. \ \mathcal{J}(\tilde{q}, \tilde{\psi}) \sim (1 + Ro_\zeta)Ro \sim 1 \times 10^{-4}, \tag{60}$$

$$c. \ \frac{E^{1/2}}{2\tilde{h}}\tilde{\nabla}\tilde{\psi} \cdot \tilde{\nabla}\tilde{q} \sim \frac{E^{1/2}}{2}(1 + Ro_\zeta)(Ro) \sim 3 \times 10^{-8}, \tag{61}$$

$$d. \ \frac{E}{2}\tilde{\nabla}^2\tilde{\zeta} \sim \frac{E}{2}Ro_\zeta \sim 1 \times 10^{-8}, \tag{62}$$

$$e. \ \frac{E^{1/2}}{2\tilde{h}}\tilde{\zeta}^2 \sim \frac{E^{1/2}}{2}Ro_\zeta^2 \sim 3 \times 10^{-6}, \tag{63}$$

$$f. \ \frac{E^{1/2}}{2\tilde{h}}\tilde{\zeta} \sim \frac{E^{1/2}}{2}Ro_\zeta \sim 3 \times 10^{-5}, \tag{64}$$

where we used $E \sim 10^{-7}$, $Ro_\zeta \sim 10^{-1}$, $Ro \sim 10^{-3}$. The two non-linear terms discussed (c. and e.) are greater or equal to the bulk viscous effects (d.) and should not be neglected in our case.

## C.3. Forcing

For now, we have introduced the forcing as an additional source of vorticity (term $\tilde{F}$ in equation (53)). The goal is to reproduce the experimental forcing such that the QG numerical model can be used as a guide and complement the experimental exploration.

In the experiment, because of the Coriolis effect, each inlet (sucking water from the tank) or outlet generates respectively a small cyclone or anticyclone right above it. This process can be modeled as a stationary source of vorticity in the form of positive or negative Gaussian sources of vorticity of radius $\ell_f$ distributed on a prescribed array. We thus define $N$ forcing points distributed over the numerical domain, and at each point, we place a Gaussian source of vorticity such that

$$\tilde{F}(\tilde{x}, \tilde{y}) = \tilde{F}_0 \sum_{i=1}^{N} (-1)^i \exp\left(-\left[\frac{\tilde{x} - \tilde{x}_i}{\tilde{\ell}_f}\right]^2 - \left[\frac{\tilde{y} - \tilde{y}_i}{\tilde{\ell}_f}\right]^2\right), \tag{65}$$

where $(\tilde{x}, \tilde{y})$ are non-dimensional cartesian coordinates, the pairs $(\tilde{x}_i, \tilde{y}_i)$, $i \in [[1, N]]$ are the centre of each forcing vortex, $\ell_f$ their radius and $F_0$ the forcing amplitude. Various forcing configurations were explored, as listed in Table 4.

### C.4. Zonal flow evolution equation

We perform a Reynolds decomposition of the flow by writing the velocity field as an azimuthal (zonal) average plus some fluctuations:

$$\langle X \rangle_\phi = \frac{1}{2\pi} \int_{\phi=0}^{2\pi} X \, d\phi, \tag{66}$$

$$\tilde{u}_\phi = \langle \tilde{u}_\phi \rangle_\phi + \tilde{u}'_\phi = \tilde{U}_\phi(\tilde{\rho}, \tilde{t}) + \tilde{u}'_\phi(\tilde{\rho}, \phi, \tilde{t}) \tag{67}$$

$$\tilde{u}_\rho = \langle \tilde{u}_\rho \rangle_\phi + \tilde{u}'_\rho = \tilde{U}_\rho(\tilde{\rho}, \tilde{t}) + \tilde{u}'_\rho(\tilde{\rho}, \phi, \tilde{t}). \tag{68}$$

Note that the zonally-averaged radial velocity, $\tilde{U}_\rho$, is not zero because of the Ekman pumping. Instead, from (47) we have

$$\tilde{U}_\rho = \frac{E_R^{1/2}}{2\tilde{h}} \tilde{U}_\phi. \tag{69}$$

Taking the zonal mean of the $\phi$-component of the momentum equation (31) gives an equation for the zonal flow evolution:

$$\frac{\partial \tilde{U}_\phi}{\partial \tilde{t}} + \tilde{U}_\rho \frac{\partial \tilde{U}_\phi}{\partial \tilde{\rho}} - \frac{\tilde{U}_\phi \tilde{U}_\rho}{\tilde{\rho}} + \tilde{U}_\rho = \underbrace{-\left\langle \tilde{u}'_\rho \frac{\partial \tilde{u}'_\phi}{\partial \tilde{\rho}} - \frac{\tilde{u}'_\phi \tilde{u}'_\rho}{\tilde{\rho}} \right\rangle_\phi}_{\mathcal{R}} + \frac{E_R}{2} \left( \tilde{\nabla}^2 \tilde{U}_\phi - \frac{\tilde{U}_\phi^2}{\tilde{\rho}^2} \right). \tag{70}$$

Equation (70) shows that the zonal flow is driven by non-linear interactions between eddies, $\mathcal{R}$. This term corresponds to the divergence of horizontal Reynolds stresses, sometimes referred to as the eddy momentum flux.

### C.5. Numerical methods

The numerical model used in this paper is directly adapted from a previous code used to study convection in a plane layer (Favier, Silvers and Proctor, 2014; Matthews, Proctor and Weiss, 1995).

The non-axisymmetric and axisymmetric motions are solved separately. For the non-axisymmetric motions (azimuthal Fourier modes $|m| > 0$), we solve the vorticity-stream-function system (53)-(56) with free-slip boundary conditions at both the outer and inner boundaries. The unknown $\tilde{\psi}$ and $\tilde{\zeta}$ are decomposed into their Fourier components up to degree $m = 2048$ and we use a pseudo-spectral method in the azimuthal direction with the 2/3$^{\mathrm{rd}}$-rule dealiasing method. In the radial direction, we use centered finite differences of fourth-order on 1024 uniformly-distributed points. The zonal mode ($m = 0$) is computed separately by solving equation (70). Time integration is performed with an implicit Crank-Nicolson scheme for the linear operator, applied directly in the Fourier space. An explicit third-order Adams Bashforth scheme is used for the non-linear terms. We use an adaptative time-step so that the CFL stability condition is verified, with a safety factor of 0.1. The code is parallelized using MPI, and we typically ran the simulations over 64 CPUs. For the simulations with $E_R = 1.25 \times 10^{-7}$, the typical computational time is of $\sim$20 hours for $\sim 3,000$ rotation times.

## D. Discrete Bessel-Fourier transform

The experimental velocity fields being obtained on a discrete grid, the Bessel-Fourier transform coefficients need to be computed using numerical quadrature, and we will thus obtain a finite set of discrete coefficients $\{\hat{a}_{nm} | n = 1, 2, \cdots, N_\rho$ and $m = -N_\phi/2 + 1, \cdots, 0, 1, 2, \cdots, N_\phi/2\}$. For simplicity, we denote $f_{ji} = f(\rho_i, \phi_j)$ where $\rho_i$ and $\phi_j$ are the discrete radial and azimuthal positions respectively ($i \in [\![1, N_\rho]\!]$, $j \in [\![1, N_\phi]\!]$).

*Discrete Fourier transform* The angular part of the transform is performed using the Matlab *fft* function. For each radius, we thus compute a discrete FFT in the azimuthal direction. The direct and inverse discrete transforms read respectively

$$\hat{f}_m(\rho) = \sum_{j=1}^{N_\phi} f_j(\rho) e^{-2i\pi m \frac{j-1}{N_\phi}}, \tag{71}$$

$$f_j(\rho) \;=\; \frac{1}{N_\phi} \sum_{m=1}^{N_\phi} \hat{f}_m(\rho) e^{2i\pi(j-1)\frac{m-1}{N_\phi}} . \tag{72}$$

The associated discrete version of the Parseval relation is

$$\sum_{j=1}^{N_\phi} |f_j|^2 = \frac{1}{N_\phi^2} \sum_{m=1}^{N_\phi} \hat{f}_m \hat{f}_m^* . \tag{73}$$

The kinetic energy per azimuthal wavenumber can then be computed as

$$\forall\, m, \quad E_m(\rho) = \frac{1}{N_\phi^2} \left[ \hat{u}_m \hat{u}_m^* + \hat{v}_m \hat{v}_m^* \right] \tag{74}$$

(see Durran, Weyn and Menchaca, 2017, for more details).

*Discrete Hankel transform* For each azimuthal mode $m$, its Fourier transform coefficient $\hat{f}_m(\rho)$ has a radial structure onto which we perform a discrete Hankel transform of order $m$. We use the Matlab algorithm provided by Guizar-Sicairos and Gutiérrez-Vega (2004). Briefly, for each mode $m$, the discrete Hankel transform coefficients are computed following

$$\forall\, m, \quad \hat{\hat{f}}_{nm} = \frac{1}{\pi V^2} \sum_{i=1}^{N_\rho} \frac{\hat{f}_{mi}}{J_{m+1}^2(\alpha_{mi})} J_m\left( \frac{\alpha_{nm}\alpha_{mi}}{S} \right), \tag{75}$$

$$\forall\, m, \quad \hat{f}_{mi} = \frac{1}{\pi R^2} \sum_{n=1}^{N_\rho} \frac{\hat{\hat{f}}_{nm}}{J_{m+1}^2(\alpha_{mn})} J_m\left( \frac{\alpha_{nm}\alpha_{mi}}{S} \right) \tag{76}$$

where $\hat{f}_{mi} = \hat{f}_m(\rho_i)$ is the radial structure of each Fourier mode, $R$ is the maximum radius, $V$ is the maximum radial wavenumber, which is different for each mode $m$ ($V_m = \alpha_{N_\rho+1,m}/(2\pi R)$), and $S = 2\pi R V$. Here again, $\alpha_{nm}$ is the $n^{\text{th}}$ zero of the Bessel function of the first kind of order $m$, $J_m$. The corresponding discrete Parseval relation is

$$\forall\, m, \quad \sum_{i=1}^{N_\rho} \frac{|\hat{f}_{mi}|^2}{2\pi^2 V^2 J_{m+1}^2(\alpha_{mi})} = \sum_{n=1}^{N_\rho} \frac{|\hat{\hat{f}}_{mn}|^2}{2\pi^2 R^2 J_{m+1}^2(\alpha_{mn})} . \tag{77}$$

Taking into account both the Fourier and Hankel transforms, the kinetic energy for each mode $(m,n)$ can then be expressed as

$$\forall\, (n,m), \quad E_{nm} = \frac{1}{N_\phi^2} \frac{1}{2\pi^2 R^3 J_{m+1}^2(\alpha_{nm})} \left[ \hat{\hat{u}}_{nm} \hat{\hat{u}}_{nm}^* + \hat{\hat{v}}_{nm} \hat{\hat{v}}_{nm}^* \right]. \tag{78}$$

Note that $\hat{\hat{u}}_{nm}$ is in $\mathrm{m^3 s^{-1}}$, and thus $E_{nm}$ is in $\mathrm{m^3 s^{-2}}$, which is homogeneous to a kinetic energy per wavenumber. The corresponding wavenumber is $k_{nm} = \alpha_{nm}/(2\pi R)$. The zonal kinetic energy spectrum $E_z$ is the spectrum of the axisymmetric mode $m = 0$ only, and the residual energy spectra $E_r$ is the sum of the contributions of the remaining modes:

$$E_z \;=\; E_{n0}, \tag{79}$$

$$E_r \;=\; \sum_{m, m \neq 0} E_{nm}. \tag{80}$$

Note that each azimuthal mode, $m$, has a different corresponding wavevector, $k_{nm}$ because the zeros of a Bessel function, $\alpha_{nm}$, depends on its order $m$. To perform the summation leading to the residual spectrum, we hence use data binning. More precisely, the magnitudes of all spectra with $m \neq 0$ are combined and sorted into wavevector bins, which are defined to correspond to the zonal wavevector $k_{n0}$. Then, the amplitudes contained in each bin are added together. This process provides a single residual spectrum associated to a unique wavevector.

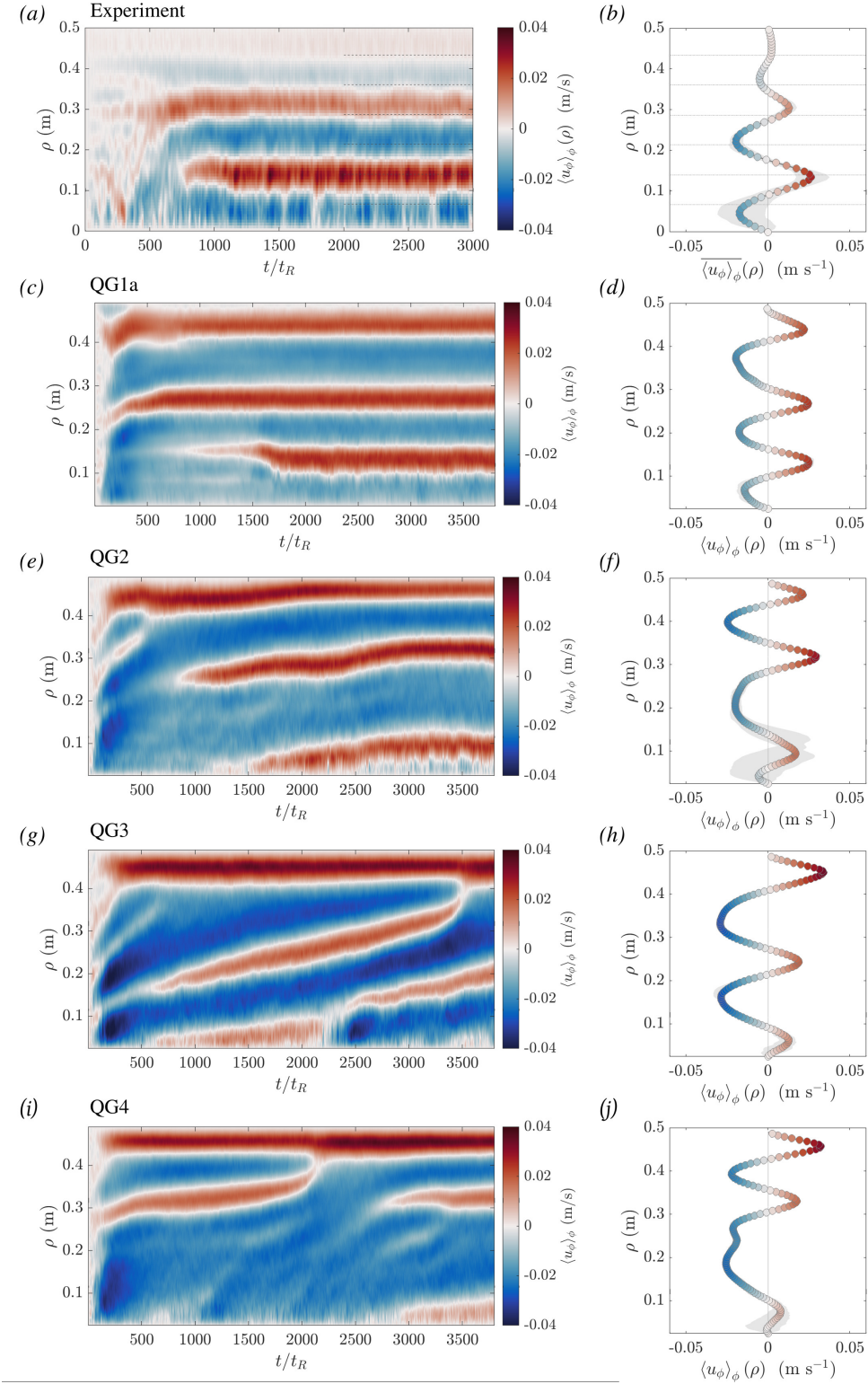

**Figure 12:** Space-time diagrams and zonal flow profiles for a typical experiment in regime II *(a,b)*, and for QG simulations QG1a *(c,d)*, QG2 *(e,f)*, QG3 *(g,h)* and QG4 *(i,j)* (see Table 4).

Table 6: Dimensional and non-dimensional parameters of present and previous experimental studies of zonal jets. $u_{rms}$ is the total root-mean-squared velocity, $h_0$ the mean fluid thickness, $R$ the tank radius, $\Omega$ the rotation rate, and the $\beta$-effect arises from the paraboloidal free surface, a sloping bottom, or both depending on the study. When a range of parameters was explored, we indicate the values of the most extreme cases. For rotating annuli, $R$ is the gap width, not the external radius. The non-dimensional parameters definitions are provided in Table 1. For the planetary flows, $u_{rms}$, $\beta$ and $R_\beta$ are taken from Table 13.1 in Galperin and Read (2019). The depth of the fluid layer for Jupiter and Saturn is taken from Galanti and Kaspi (2021). The kinematic viscosity used to compute the Ekman and Reynolds numbers is assumed to be $\nu \sim 3 \times 10^{-7}$ m² s⁻¹ for Jupiter and Saturn, $1.5 \times 10^{-5}$ m² s⁻¹ (Gastine et al., 2014), $1.5 \times 10^{-5}$ m² s⁻¹ for the atmosphere, and $1 \times 10^{-6}$ m² s⁻¹ for the oceans.

| Ref. | Forcing | $u_{rms}$ (m s⁻¹) | $h_0$ (m) | $R$ (m) | $\Omega$ (RPM) | $\beta$ (m⁻¹ s⁻¹) | $E$ | $Re$ | $Ro = Re \times E$ | $R_\beta^* = 0.5 R_\beta^S$ |
|---|---|---|---|---|---|---|---|---|---|---|
| Present | Sinks & sources ($L_f \sim 7$cm) | $5 \times 10^{-3}$ | 0.58 | 0.49 | 75 | 50 | $3.78 \times 10^{-7}$ | 19720 | $7.46 \times 10^{-3}$ | 2.70 |
| Burin, Caspary, Edlund, Ezeta, Gilson, Jr, McNulty, Squire and Tynan (2019) | Sinks & sources ($L_f \sim 6.5$cm) | $2.50 \times 10^{-2}$ | 0.22 | 0.13 | 100 | 18 | $1.97 \times 10^{-6}$ | 5500 | $1.09 \times 10^{-2}$ | 1.91 |
| Cabanes et al. (2017) | Sinks & sources ($L_f \sim 10$cm) | $5 \times 10^{-2}$ | 0.5 | 0.5 | 75 | 74 | $5.09 \times 10^{-7}$ | 17500 | $8.91 \times 10^{-3}$ | 2.72 |
| Read, Jacoby, Rogberg, Wordsworth, Yamazaki, Miki-Yamazaki, Young, Sommeria, Didelle and Vibaud (2015) | Barotropic thermal conv. | $4.00 \times 10^{-3}$ | 0.8 | 7.5 | 22 | $6.20 \times 10^{-2}$ | $9.95 \times 10^{-6}$ | 3200 | $3.18 \times 10^{-2}$ | 1.78 |
| Zhang and Afanasyev (2014) | Electromag. ($L_f \sim 4.6$cm) | $1.40 \times 10^{-2}$ | 0.08 | 0.55 | 22 | 35 | $6.78 \times 10^{-6}$ | 1120 | $7.60 \times 10^{-2}$ | 1.83 |
| Smith, Speer and Griffiths (2014) | Diff-heated rotating annulus | $3.00 \times 10^{-3}$ | 0.15 | 0.49 | 38 | 107 | $1.12 \times 10^{-5}$ | 450 | $5.03 \times 10^{-2}$ | 1.89 |
| Di Nitto, Espa and Cenedese (2013) | Electromag ($L_f \sim 1.2$cm) | $8.00 \times 10^{-3}$ | 0.03 | 0.5 | 32 | 35 | $3.32 \times 10^{-4}$ | 240 | $7.96 \times 10^{-2}$ | 1.38 |
| Afanasyev and Craig (2013) | Electromag ($L_f \sim 4.6$cm) | $5.80 \times 10^{-3}$ | 0.1 | 0.55 | 35 | 35 | $4.34 \times 10^{-5}$ | 580 | $2.52 \times 10^{-2}$ | 1.76 |
| Afanasyev, O'Leary, Rhines and Lindahl (2012) | Local buoyancy source | $5.00 \times 10^{-3}$ | 0.12 | 0.65 | 6 | 6 | $3.01 \times 10^{-5}$ | 600 | $1.81 \times 10^{-2}$ | 1.51 |
| Espa, Bordi, Frisius, Foodrich, Cenedese and Sutera (2012) | Electromag ($L_f \sim 2$cm) | $1.10 \times 10^{-2}$ | 0.01 | 0.18 | 24 | 44 | $3.98 \times 10^{-3}$ | 110 | $4.38 \times 10^{-1}$ | 1.20 |
| Wordsworth, Read and Yamazaki (2008) | Diff-heated rotating annulus | $1.80 \times 10^{-2}$ | 0.22 | 0.1 | 37 | 29 | $5.33 \times 10^{-5}$ | 3960 | $2.11 \times 10^{-2}$ | 2.15 |
| Read, Yamazaki, Lewis, Williams, Wordsworth, Miki-Yamazaki, Sommeria and Didelle (2007) | Spray of dense water | $3.40 \times 10^{-3}$ | 0.55 | 7.5 | 1.5 | 8 | $2.10 \times 10^{-5}$ | 1870 | $3.94 \times 10^{-2}$ | 1.50 |
| Afanasyev and Wells (2005) | Electromag ($L_f \sim 2$cm) | $5.00 \times 10^{-3}$ | $5.00 \times 10^{-3}$ | 0.15 | 14 | 40 | $2.73 \times 10^{-5}$ | 25 | $6.82 \times 10^{-3}$ | 1.00 |
| Aubert, Jung and Swinney (2002) | Sinks & sources ($L_f \sim 14$cm) | $2.00 \times 10^{-4}$ | 0.19 | 0.32 | 17 | 17 | $1.76 \times 10^{-4}$ | 3800 | $6.70 \times 10^{-3}$ | 1.74 |
| Bastin and Read (1998) | Diff-heated rotating annulus | $3.00 \times 10^{-3}$ | 0.14 | 0.06 | 43 | 91 | $1.13 \times 10^{-5}$ | 420 | $4.76 \times 10^{-3}$ | 1.81 |
| **Planet** | | $u_{rms}$ (m s⁻¹) | $h_0$ (m) | $R$ (m) | $\Omega$ (rad/s) | $\beta$ (m⁻¹ s⁻¹) | $E$ | $Re$ | $Ro = Re \times E$ | $R_\beta^S$ |
| Jupiter | – | 50 | $2 \times 10^6$ | $7 \times 10^7$ | $7 \times 10^{-4}$ | $3 \times 10^{-12}$ | $4 \times 10^{-16}$ | $3 \times 10^4$ | $1 \times 10^{-11}$ | 5 |
| Saturn | – | 40 | $7 \times 10^6$ | $6 \times 10^7$ | $6 \times 10^{-4}$ | $3 \times 10^{-12}$ | $3 \times 10^{-17}$ | $9 \times 10^4$ | $3 \times 10^{-12}$ | 5.3 |
| Earth's atmosphere | – | 4 | $1 \times 10^4$ | $6.4 \times 10^6$ | $2 \times 10^{-3}$ | $2 \times 10^{-11}$ | $2 \times 10^{-9}$ | $3 \times 10^8$ | 6 | 1.6 |
| Earth's oceans | – | 0.1 | $1 \times 10^3$ | $6.4 \times 10^6$ | $2 \times 10^{-3}$ | $1 \times 10^{-11}$ | $1 \times 10^{-8}$ | $1 \times 10^6$ | 1 | 1.4 |